\documentclass[]{aastex631}
\usepackage{chngcntr}
\submitjournal{ApJL}

\shorttitle{M-Dwarf CO/O$_2$ Runaway Revisited}
\shortauthors{Ranjan et al.}
\begin{document}

\title{The Importance of the Upper Atmosphere to CO/O$_2$ Runaway on Habitable Planets Orbiting Low-Mass Stars}

\correspondingauthor{Sukrit Ranjan}
\email{sukrit@arizona.edu}

\author[0000-0002-5147-9053]{Sukrit Ranjan}
\altaffiliation{Center for Interdisciplinary Exploration and Research in Astrophysics \& Department of Physics and Astronomy, Northwestern University, Evanston, 60201, USA}
\altaffiliation{Blue Marble Space Institute of Science, Seattle, 98104, USA}
\affiliation{University of Arizona, Lunar and Planetary Laboratory/Department of Planetary Sciences, Tucson, 85721, USA}

\author[0000-0002-2949-2163]{Edward W. Schwieterman}
\altaffiliation{Blue Marble Space Institute of Science, Seattle, 98104, USA}
\affiliation{Department of Earth and Planetary Sciences,
University of California at Riverside, Riverside, 92521, USA}

\author[0000-0003-1906-5093]{Michaela Leung}
\affiliation{Department of Earth and Planetary Sciences,
University of California at Riverside, Riverside, 92521, USA}

\author[0000-0003-2281-1990]{Chester E. Harman}
\affiliation{Planetary Systems Branch, Space Science and Astrobiology Division, NASA Ames Research Center, Moffett Field, 94035, USA}%\\
%1667 K Street NW, Suite 800 \\

\author[0000-0003-2215-8485]{Renyu Hu}
\affiliation{Jet Propulsion Laboratory, California Institute of Technology, Pasadena, 91109, USA}
\affiliation{Division of Geological and Planetary Sciences, California Institute of Technology, Pasadena, 91125, USA}%



\begin{abstract}

%250 word limit.
Efforts to spectrally characterize the atmospheric compositions of temperate terrestrial exoplanets orbiting M-dwarf stars with the James Webb Space Telescope (JWST) are now underway. Key molecular targets of such searches include O$_2$ and CO, which are potential indicators of life. Recently, it was proposed that CO$_2$ photolysis generates abundant ($\gtrsim0.1$ bar) abiotic O$_2$ and CO in the atmospheres of habitable M-dwarf planets with CO$_2$-rich atmospheres, constituting a strong false positive for O$_2$ as a biosignature and further complicating efforts to use CO as a diagnostic of surface biology. Importantly, this implied that TRAPPIST-1e and TRAPPIST-1f, now under observation with JWST, would abiotically accumulate abundant  O$_2$ and CO, if habitable. Here, we use a multi-model approach to re-examine photochemical O$_2$ and CO accumulation on planets orbiting M-dwarf stars. We show that photochemical O$_2$ remains a trace gas on habitable CO$_2$-rich M-dwarf planets, with earlier predictions of abundant O$_2$ and CO due to an atmospheric model top that was too low to accurately resolve the unusually-high CO$_2$ photolysis peak on such worlds.  Our work strengthens the case for O$_2$ as a biosignature gas, and affirms the importance of CO as a diagnostic of photochemical O$_2$ production.  However, observationally relevant false positive potential remains, especially for O$_2$'s photochemical product O$_3$, and further work is required to confidently understand O$_2$ and O$_3$ as biosignature gases on M-dwarf planets. 
\end{abstract}

%This artifact arose because the unique photochemical regime unveiled on M-dwarf planets results in much higher CO$_2$ photolysis altitudes, illustrating the challenges of projecting models calibrated in the solar system regime to the novel parameter space accessible on exoplanets. 


\section{Introduction} \label{sec:intro}
The launch of the James Webb Space Telescope (JWST) has begun a new era in the characterization of exoplanet atmospheres. JWST has already revolutionized the study of gas giant exoplanet atmospheres, detecting novel chemical species and using them to infer operant chemical processes \citep{Tsai2023}. In future, JWST will similarly seek to characterize the atmospheres of smaller exoplanets including potentially-habitable temperate terrestrial worlds, and such observations have already begun \citep{Lustig-Yaeger2019, Lafreniere2017_JWST_T1d, JWST_T1e_Lewis, Lim2021_JWST, Stevenson2021_JWST}. Such observations may constrain surface processes such as volcanism, geochemical cycling, and the presence of life \citep{Kaltenegger2010, Kaltenegger2010b, Misra2015, Krissansen-Totton2018biosig, Lehmer2020, Fauchez2020, Zhan2021, Ranjan2022b}.

Particularly relevant in the quest to characterize temperate terrestrial exoplanets are planets with anoxic, CO$_2$-rich atmospheres orbiting M-dwarf stars. CO$_2$-rich atmospheres are expected to be ubiquitous on terrestrial exoplanets, as they are in the Solar System, thanks to robust outgassing of CO$_2$ by basaltic magmatism \citep{Gaillard2014, CatlingKasting2017}. Such high mean-molecular-mass secondary atmospheres are only accessible to atmospheric characterization with JWST for planets orbiting M-dwarf stars. Planets with CO$_2$-rich atmospheres orbiting M-dwarf stars are therefore highly observationally relevant, and it is critical to understand their atmospheric photochemistry to accurately interpret anticipated observations of their atmospheres \citep{Shields2016rev, Catling2018}.

Atmospheric photochemical models of planets with CO$_2$-rich atmospheres orbiting M-dwarf stars are historically discrepant, with observationally-relevant implications for the abundance of spectroscopically-active trace gases proposed as probes of surface processes. Disagreement has been particularly intense about photochemical CO and O$_2$ abundance in such atmospheres. While all models predicted accumulation of photochemical CO and O$_2$ on CO$_2$-rich planets orbiting M-dwarf stars, the specific predicted concentrations of CO and O$_2$ varied by many orders of magnitudes between models run on identical planetary scenarios \citep{Harman2015, Harman2018}. This is problematic, because atmospheric O$_2$ and CO are proposed as remote probes of surface processes, including the presence or absence of life \citep{Meadows2018, Schwieterman2019, Wogan2020}. Uncertainty about the efficiency of photochemical generation of these gases neuters their value as probes of surface processes. 
 
Recognizing the importance of this problem, extensive efforts have been invested to improve understanding of photochemical CO and O$_2$ on CO$_2$-rich planets orbiting M-dwarf stars. Initial efforts to simulate high-CO$_2$ atmospheres on planets orbiting M-dwarf stars resulted disagreement of up to 4 orders of magnitude in predicted pCO and pO$_2$ \citep{Tian2014, Domagal-Goldman2014, Harman2015}. \citet{Harman2015} showed some of the inter-model disagreement was due to different lower boundary conditions. \citet{Harman2018} reconciled most of the remaining inter-model disagreement by showing that it was due to different assumptions regarding the photochemical effects of lightning, which generates NO$_X$ species in a CO$_2$-N$_2$ atmosphere. These NO$_X$ species can drive catalytic chemistry which recombine CO and O$_2$ to CO$_2$. Upon including a lightning rate corresponding to modern Earth and its production of catalytic NO$_X$ species, \citet{Harman2018} found photochemical O$_2$ to be low (pO$_2<2\times10^{-4}$ bar, below the false-positive threshold) across all models and planetary scenarios they considered. 

Most recently, the possibility of abundant photochemical CO and O$_2$ was examined by \citet{Hu2020}, who highlighted that previous work did not include the key NO$_X$ reservoir species HNO$_4$ and N$_2$O$_5$. These species are relatively stable under M-dwarf UV irradiation and serve to shuttle NO$_X$ into the ocean due to their high solubility, reducing the efficacy of the NO$_X$-driven recombinative cycles. \citet{Hu2020} reproduced the results of \citet{Harman2018} for 0.05 bar CO$_2$ when excluding HNO$_4$ and N$_2$O$_5$, but found that inclusion of HNO$_4$ and N$_2$O$_5$ into the photochemical network led to generation of abundant photochemical CO and O$_2$, even assuming Earth-like lightning and its concomitant production of catalytic NO$_X$ species. This mechanism is minimally affected by updated CO$_2$ and H$_2$O NUV cross-sections \citep{Ranjan2020}, because M-dwarf NUV emission is very low, rendering the photochemical effect of the updated cross-sections modest for planets orbiting low-mass M-dwarf stars (Broussard et al. \emph{in prep}). Applying this finding to the TRAPPIST-1 system, \citet{Hu2020} found that if TRAPPIST-1e and TRAPPIST-1f had enough CO$_2$ to be globally habitable, then they would necessarily accumulate abundant ($\gtrsim0.1$ bar, and possibly $\geq1$ bar) CO and O$_2$, with high O$_3$ as well. This atmospheric state (``CO/O$_2$ runaway'') would constitute an observable false positive for O$_2$ as a biosignature, would likely constitute a false positive for O$_3$ as a biosignature, and would further complicate efforts to use CO as a diagnostic of surface biology. Importantly, this false positive would persist in comparative planetological approaches to O$_2$ as a biosignature in the TRAPPIST-1 system, since abundant photochemical O$_2$ would specifically co-occur with habitable conditions on the outer planets. With observations of the TRAPPIST-1 system already underway, it is critical to confirm the prospects for the photochemical accumulation of CO and O$_2$ on the TRAPPIST-1 planets, to understand how to interpret potential detections of O$_2$, O$_3$, and CO. 

Here, we explore further the possibility of substantial accumulation of CO and O$_2$ on CO$_2$-rich conventionally habitable planets orbiting M-dwarf stars. By ``conventionally habitable", we refer to rocky planets with atmospheres and stable surface liquid water oceans which are the prime targets for exoplanet biosignature search \citep{Kopparapu2013hz, Kasting2014PNAS}, as opposed to desiccated rocky planets, moist greenhouse rocky planets, Hycean worlds, or mini-Neptune aerial biospheres \citep{Abe2011, Gao2015, Kasting1988, Madhusudhan2021, Glidden2022}. Our basic goal is to confirm whether it is possible for the the photochemical decomposition of CO$_2$ to CO and O$_2$ to be favored in conventionally habitable planet atmospheres, even with a modern Earth-like lightning rate \citep{Harman2018, Hu2020}. We place specific emphasis on O$_2$, which is strongly motivated by the Solar System as a biosignature gas \citep{Sagan1993}, and on the TRAPPIST-1 system, now under observation by JWST. However, our findings are relevant to multiple gases and to planets orbiting M-dwarf stars in general. We reproduce the CO/O$_2$ runaway state in two independently developed photochemical models and determine its key controls. Our multi-model approach is critical because it enables us to be confident that our findings are not due to numerical or implementation errors of individual models. We describe our methods in Section~\ref{sec:methods}, report and discuss our results in Sections~\ref{sec:results} and \ref{sec:discuss}, and summarize in Section~\ref{sec:conc}. We provide additional background, supporting information, and methodological details in Appendix~\ref{sec:ap_background}-\ref{sec:ap_detailed_scenario}. 

\section{Methods \label{sec:methods}}
We employ two independent 1D photochemical models in our exploration of CO/O$_2$ runaway: the MIT Exoplanet Atmospheric Chemistry Model (MEAC; \citealt{Hu2012}) and Atmos \citep{Arney2016, lincowski2018evolved}. These models implement conceptually similar chemistry, physics, and numerical schemes, but were developed independently and share no codebase. This means that results confirmed using both models are likely to be robust to implementation errors, though they may still suffer from common-mode errors in basic physico-chemical understanding  \citep{Wen1989}. This intercomparative approach has proved useful in past studies employing exoplanet photochemical models, which are extremely complex, often nonlinear, and feature numerous free parameters \citep{Harman2015, Harman2018, Ranjan2020}. In this work, we employ MEAC and Atmos in a highly focused investigation of CO/O$_2$ runaway. A more exhaustive intercomparison engaging a much broader diversity of models in a much broader range of planetary scenarios is underway in the form of the Photochemical model Intercomparison for Exoplanets (PIE) project (PI: C. E. Harman) as part of the CUISINES initiative\footnote{\url{https://nexss.info/cuisines/}}.

\subsection{MEAC Model \& Configuration.}
We employ MEAC as updated by \citet{Ranjan2022b}, which is modified from the original \citet{Hu2012} version by elimination of errors and use of updated H$_2$O cross-sections \citep{Ranjan2020, Hu2021}. Importantly, this model is similar to the version of MEAC employed by \citet{Hu2020}, which motivated the present work. MEAC is a 1D photochemical model that calculates the steady-state vertical trace gas composition of a planetary atmosphere given temperature-pressure and eddy diffusion (vertical transport) profiles, stellar irradiation, chemical network, and chemical boundary conditions. MEAC encodes processes including photolysis (computed via delta two-stream approximation), eddy diffusion and molecular diffusion of H and H$_2$, surface emission and wet and dry deposition of chemical species, diffusion-limited escape of H and H$_2$, and formation and deposition of S$_8$ and H$_2$SO$_4$ aerosols. MEAC does not account for formation of organic haze, expected at elevated CH$_4$/CO$_2$ ratios \citep{DeWitt2009, Arney2016}; we do not explore this regime in our work. We assign a high deposition velocity of $1\times10^{-5}$ cm s$^{-1}$ for C$_2$H$_6$ to account for this omission in an ad-hoc fashion following \citet{Hu2012}. For our full CHOSN chemical network, we consider 86 species linked by 734 reactions, corresponding to the full chemistry of \citet{Hu2012} excluding the higher hydrocarbons. Appendix~\ref{sec:ap_detailed_scenario} summarizes further MEAC model set-up. 

\subsection{Atmos Model \& Configuration}
Atmos is a 1D photochemical-climate model first developed by \cite{Kasting1979-xd} and recently modified by numerous workers including \cite{Arney2016, lincowski2018evolved, teal2022effects, Leung2022}. For this work, we use the publicly available version of the code\footnote{https://github.com/VirtualPlanetaryLaboratory/atmos} with modifications as suggested by \cite{Ranjan2020}. This code has 79 species with 397 reactions, including 65 photolysis reactions. Atmos also incorporates eddy diffusion, wet and dry deposition and sulfur aerosols to simulate terrestrial environments. While not explored in this work, the model has previously been used to simulated hazy Archean-like atmospheres and has an extensive history of use for other terrestrial exoplanet applications (e.g., \citealt{Domagal-Goldman2011, Arney2016, Arney2018, Schwieterman2019, teal2022effects, Leung2022}).  

\subsection{Planetary Scenario}
We apply our photochemical models to a planetary scenario closely corresponding to the CO$_2$-dominated abiotic Earth-like planet benchmark scenario detailed in \citet{Hu2012}. We assume atmospheric structure from \citet{Hu2012} ($T(z)$, $P(z)$, $K_{z}(z)$; Figure~\ref{fig:t_p_eddy}). To emphasize enforcement of mass balance, we avoid fixed mixing ratio boundary conditions in favor of fixed surface flux and dry deposition velocity boundary conditions. We adopt surface emission of SO$_2$, H$_2$S, CH$_4$, and H$_2$ broadly consistent with Earth-like volcanism, surface production of CO and NO broadly consistent with Earth-like lightning, and generally low dry deposition velocities to simulate an abiotic planet \citep{Hu2012, Harman2018, Hu2019} (Table~\ref{tab:species_bcs}). We set wet deposition of H$_2$, O$_2$, CO, CH$_4$, C$_2$H$_2$, C$_2$H$_4$, C$_2$H$_6$, and NH$_3$ to zero, to simulate saturation on a planet with abiotic oceans \citep{Hu2012}. For stellar irradiation, we adopt the TRAPPIST-1 model 1A spectrum of \citet{Peacock2019a}.  Further details on the planet scenario are presented in Appendix~\ref{sec:ap_detailed_scenario}. 

The high pCO$_2$=0.9 bar in this scenario puts it well into the CO/O$_2$ runaway regime identified by \citet{Hu2020}. We generally avoid intermediate pCO$_2$ corresponding to the runaway transition itself, where sensitivity to photochemical assumptions is enhanced \citep{Ranjan2022b}. This choice ensures that our modeling reflects the fundamental cause of the CO/O$_2$ runaway, as opposed to ``red herrings" which have enhanced importance solely in the runaway transition regime, but do not control the overall runaway phenomenon.

\subsection{Photochemical Model Deployment}
We perform 16 targeted simulations with a combination of MEAC and Atmos to test model and parameter sensitivities and explore CO/O$_{2}$ runaway (Table~\ref{tab:pco_po2_modelruns}). This includes the stellar spectrum (including the Sun, TRAPPIST-1, and GJ 876), the partial pressure of CO$_{2}$ (pCO$_{2}$), the chemical network: the full CHOSN network as described for both models, then chemical networks without N$_2$O$_5$ and HNO$_4$ (CHOSN - (N$_2$O$_5$+HNO$_4$)), without nitrogen chemistry (CHOS), and without nitrogen and sulfur chemistry (CHO). We additionally test the impact of the shortwave UV cutoff in the stellar spectrum ($\lambda_{crit}$), the maximum altitude of the model ($z_{max}$) and model grid resolution ($\Delta z$), and the magnitude of vertical transport (scaled by a factor $K_{scale}$). The results of these tests are summarized in the last two columns of Table~\ref{tab:pco_po2_modelruns} in which we show the surface-level CO and O$_{2}$ partial pressures (pCO and pO$_{2}$, respectively). 

\subsection{Synthetic Spectra\label{sec:methods_synspec}}
To illustrate the observational implications of our photochemical simulations, we simulate planetary transmission spectra. We simulate these synthetic spectra using the Planetary Spectrum Generator (PSG; \citealt{Villanueva2018, Villanueva2022}) by adapting the example presented at \url{https://github.com/nasapsg/globes/blob/main/atmos.py}. PSG is a widely used community tool for simulating exoplanet transmission spectra (e.g., \citealt{Suissa2020, Fauchez2020, Pidhorodetska2020}). We choose planetary and stellar parameters consistent with our photochemical simulations (i.e. 1 M$_\earth$, 1 $R_\earth$ planet orbiting TRAPPIST-1 like star). We include absorption due to H$_2$O, CH$_4$, C$_2$H$_6$, CO$_2$, O$_2$, O$_3$, CO, H$_2$CO, NO, NO$_2$, SO$_2$, N$_2$O, and N$_2$, as well as the effects of Rayleigh scattering, refraction and all collision-induced absorption (CIA) included in PSG at a resolution of $R=500$. PSG reports spectral fractional transit depths $d(\lambda)$, where
\begin{equation}
    d(\lambda)=(\frac{R_P + z_{atm}(\lambda)}{R_\star})^2\label{eqn:spectrandepth}
\end{equation}
\noindent where $R_P$ is the solid radius of the planet, $R_\star$ is the radius of the star, and $z_{atm}(\lambda)$ is the wavelength-dependent effective height of the atmosphere, which is specific to a given host star. To obtain a more general metric of the transmission spectrum that is not host star-specific and can be compared to other simulated transmission spectra, we also compute the effective height of the atmosphere $z_{atm}(\lambda)$ by solving Equation~\ref{eqn:spectrandepth} for $z_{atm}(\lambda)$.

\section{Results \label{sec:results}}
\subsection{Photochemistry}
We find that the occurrence of CO/O$_2$ runaway for conventionally habitable planets with high-CO$_2$ atmospheres is primarily controlled by the resolution of the CO$_2$ photolysis peak, via choice of $z_{max}$ and $\lambda_{crit}$.

We find that the details of the photochemical network are not the primary control for CO/O$_2$ runaway. Previously, it was reported that inclusion of N$_2$O$_5$ and HNO$_4$ into photochemical networks resulted in CO/O$_2$ runaway for CO$_2$-dominated atmospheres orbiting M-dwarf stars, because the reactive nitrogen which could catalyze the recombination of CO and O$_2$ back to CO$_2$ was instead converted into HNO$_4$ and N$_2$O$_5$ and sequestered into the ocean. Indeed, CO/O$_2$ runaway is sensitive to the details of the chemical network at intermediate pCO$_2$, corresponding to the onset of runaway (Table~\ref{tab:pco_po2_modelruns}, Run 8-9), as originally reported \citep{Hu2020}. In this intermediate atmospheric regime, sensitivity to photochemical details is enhanced, because the dominant forcing on the atmosphere is changing \citep{Ranjan2022b}. However, our simulations show that for CO$_2$-rich atmospheres which are firmly in the runaway regime, the runaway persists when N$_2$O$_5$ and HNO$_4$ are removed from the photochemical network (Table~\ref{tab:pco_po2_modelruns}, Run 1-2). Indeed, it persists even when nitrogen and sulfur chemistry are excised entirely from the network (Table~\ref{tab:pco_po2_modelruns}, Run 3-4). This demonstrates that the details of the photochemical network are not the fundamental driver of CO/O$_2$ runaway.

\begin{deluxetable}{ccccccccc|cc}
\tablecaption{Predicted pCO and pO$_2$ for our baseline 0.9 bar CO$_2$, 0.1 bar N$_2$ planetary scenario for different model assumptions. We have \textbf{bolded} instances of CO/O$_2$ runaway, which we define on an ad-hoc basis as pCO$>0.1$ bar and pO$_2>0.1$ bar. Assumptions regarding the details of the chemical scheme are not the main control on CO/O$_2$ runaway. Rather, assumptions regarding the shortwave UV cutoff and especially regarding the model top are the main control on CO/O$_2$ runaway, through their influence on resolution of the CO$_2$ photolysis peak. This finding is multi-model. \label{tab:pco_po2_modelruns}}
\tablewidth{0pt}
\tablehead{\colhead{Run} & \colhead{Model} & \colhead{Star} & \colhead{pCO$_2$} & \colhead{Network} & \colhead{$\lambda_{crit}$} & \colhead{$z_{max}$} & \colhead{$\Delta z$} & \colhead{$K_{scale}$} &  \colhead{pCO} & \colhead{pO$_2$}\\
& & & (bar) & & (nm) & (km) & (km) & & (bar) & (bar)
}
\startdata
 0 & MEAC & Sun & 0.9 & CHOSN & 110 & 54 & 1 & 1 & $5\times10^{-6}$ & $2\times10^{-20}$\\
1 & MEAC & TRAPPIST-1 & 0.9 & CHOSN & 110 & 54 & 1 & 1 & \textbf{0.8} & \textbf{1}\\
2 & MEAC & TRAPPIST-1 & 0.9 & CHOSN - (N$_2$O$_5$+HNO$_4$) & 110 & 54 & 1 & 1 & \textbf{0.8} & \textbf{1}\\
3 & MEAC & TRAPPIST-1 & 0.9 & CHOS & 110 & 54 & 1 & 1 & \textbf{0.8} & \textbf{1}\\
4 & MEAC & TRAPPIST-1 & 0.9 & CHO & 110 & 54 & 1 & 1 & \textbf{0.8} & \textbf{1}\\
5 & MEAC & TRAPPIST-1 & 0.9 & CHO & 110 & 100 & 1 & 1 & 0.07 & $1\times10^{-5}$\\
6 & MEAC & TRAPPIST-1 & 0.9 & CHO & 120 & 54 & 1 & 1 & 0.07 & $3\times10^{-4}$\\
7 & MEAC & TRAPPIST-1 & 0.9 & CHO & 120 & 100 & 1 & 1 & 0.07 & $1\times10^{-5}$ \\
8 & Atmos & TRAPPIST-1 & 0.9 & CHOSN & 117.6 & 100 & 0.5 & 1 & 0.095 & $1.9\times10^{-4}$ \\
9 & Atmos & TRAPPIST-1 & 0.9 & CHOSN & 117.6 & 55 & 0.5 & 1 & \textbf{1.65} & \textbf{0.13}\\
10 & MEAC & GJ 876 & 0.05 & CHOSN & 110 & 54 & 1 & 1 & 0.06 & $2\times10^{-3}$\\
11 & MEAC & GJ 876 & 0.05 & CHOSN - (N$_2$O$_5$+HNO$_4$)& 110 & 54 & 1 & 1 & 0.03 & $5\times10^{-9}$ \\
12 & MEAC & TRAPPIST-1 & 0.9 & CHO & 110 & 54 & 0.1 & 1 & 0.08 & $7\times10^{-4}$\\
13 & MEAC & TRAPPIST-1 & 0.9 & CHO & 110 & 54 & 1 & $10^{3}$ & 0.26 & $2\times10^{-4}$\\
14 & MEAC & TRAPPIST-1 & 0.9 & CHO & 110 & 100 & 1 & $10^{-3}$ & $5\times10^{-8}$ & $2\times10^{-19}$\\
15 & MEAC & TRAPPIST-1 & 0.9 & CHO & 110 & 100 & 5 & 1 & 0.01 & $7\times10^{-11}$\\
16 & MEAC & TRAPPIST-1 & 0.9 & CHO & 110 & 100 & 0.2 & 1 & 0.04 & $2\times10^{-5}$\\
\enddata
%\tablecomments{Text} 
\end{deluxetable}

Instead, CO/O$_2$ runaway for high-CO$_2$ atmospheres is driven by choice of $z_{max}$ and $\lambda_{crit}$. Specifically, we found that either increasing $z_{max}$ from 54 km to 100 km or increasing $\lambda_{crit}$ from 110 nm to 120 nm was adequate to inhibit the atmosphere from becoming CO/O$_2$ dominated, though concentrations of both gases remained high relative to the Sun-like star case (Table~\ref{tab:pco_po2_modelruns}, Run 0, 5-6). Both a low $\lambda_{crit}$ and a low $z_{max}$ were required to induce a runaway. Critically, this finding is multi-model: by adjusting $z_{max}$ from 100 km to 55 km, we were able to induce CO/O$_2$ runaway in the independently-developed Atmos photochemical model\footnote{\citealt{Arney2016}; main branch, Archean template, commit b98a47527d14e6cf82e0edf4640fc7fba6d09cc8} as well, despite its omission of HNO$_4$ and N$_2$O$_5$ (Table~\ref{tab:pco_po2_modelruns}, Run 10-11). This confirms that $\lambda_{crit}$ and $z_{max}$, and not the details of the photochemical network, are the main drivers of model predictions of CO/O$_2$ runaway. 

To elucidate the mechanism by which $\lambda_{crit}$ and $z_{max}$ drive CO/O$_2$ runaway, we modeled atmospheres with pCO$_2$=0.01-0.1 bar (pN$_2$=0.99-0.9 bar). We explored both truncated atmospheres (model tops set at 0.34 $\mu$bar, as in the baseline 54 km case) and extended atmospheres (model tops set at 100 km; 4 nanobar for pCO$_2=0.1$ bar). We chose this pCO$_2$ range because it corresponded to the onset of runaway in the truncated atmospheres, enabling us to study the numerical drivers of this phenomenon, and because at higher pCO$_{2}$, photochemical O$_2$/O$_{3}$ layers emerge in the truncated atmospheres, which strongly suppresses tropospheric near-UV (NUV) radiation and makes it hard to study the subtle changes in CO$_2$ and H$_2$O photolysis rates that ultimately drive the runaway phenomenon. 

\begin{figure}[h]
\centering
\includegraphics[width=0.8\linewidth]{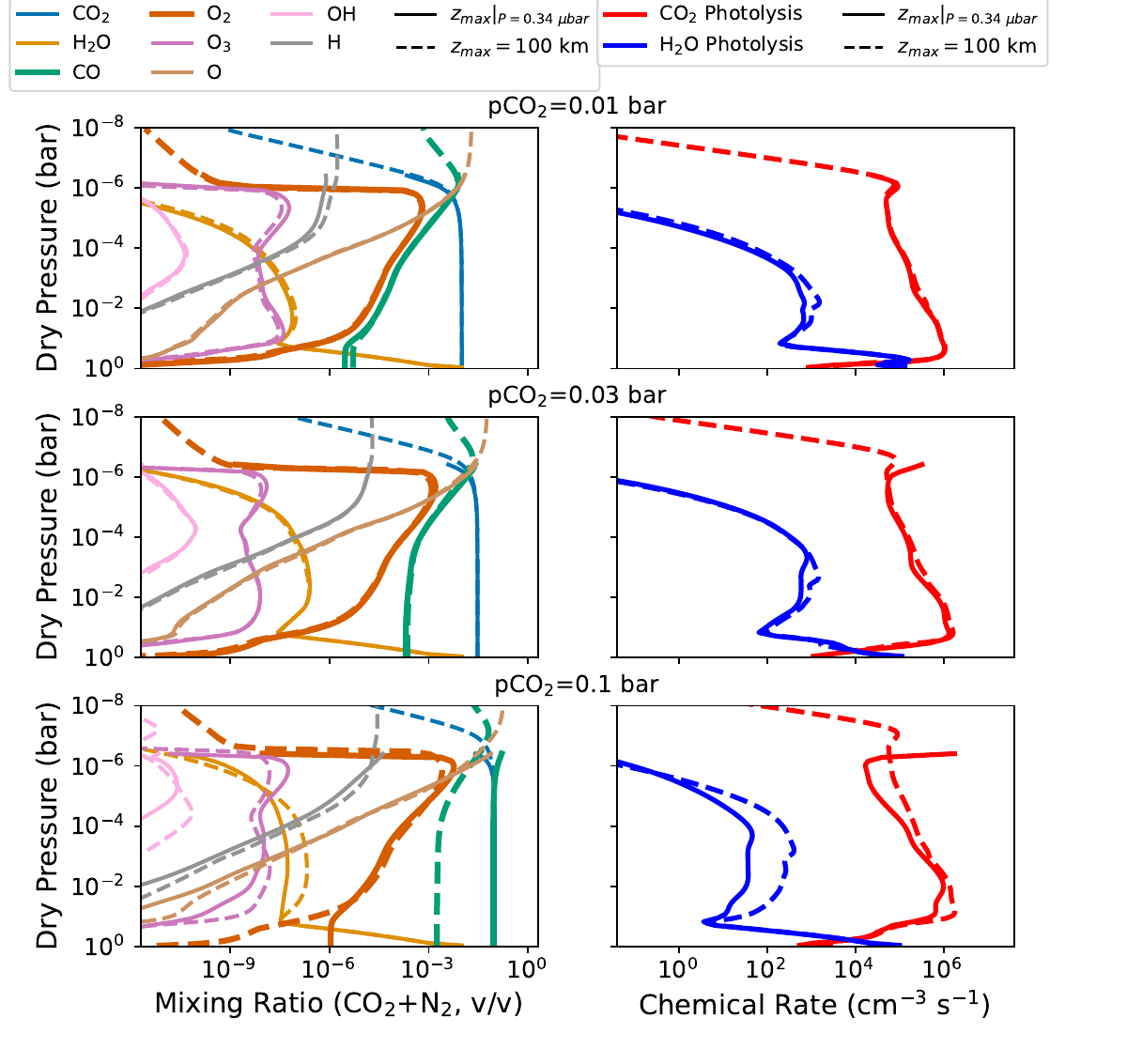}
\caption{Non-resolution of the CO$_2$ photolysis peak drives CO/O$_2$ runaway. Concentrations of key atmospheric species (left column) and CO$_2$ and H$_2$O photolysis rates (right column) as a function of altitude, for pCO$_2$=0.01, 0.03, and 0.1 bar, CHO chemistry only (S, N excluded). The key photochemical products CO and O$_2$ are highlighted with thicker lines in the left column. ``Dry pressure" refers to the atmospheric pressure due to CO$_2$ and N$_2$,  excluding the contribution from water vapor. Solid lines correspond to calculations with model top at $0.34~\mu$bar, while dashed lines correspond to calculations with a higher 100 km model top. At 0.01 bar, the photolysis peak is resolved by both models, and they report identical results. As pCO$_2$ increases, the CO$_2$ photodissociation peak moves upward and is not resolved by the  calculations with $0.34~\mu$bar model top, resulting in large increases in pCO and especially pO$_2$.} 
\label{fig:photolysis_peak}
\end{figure}

\begin{figure}[h]
\centering
\includegraphics[width=0.8\linewidth]{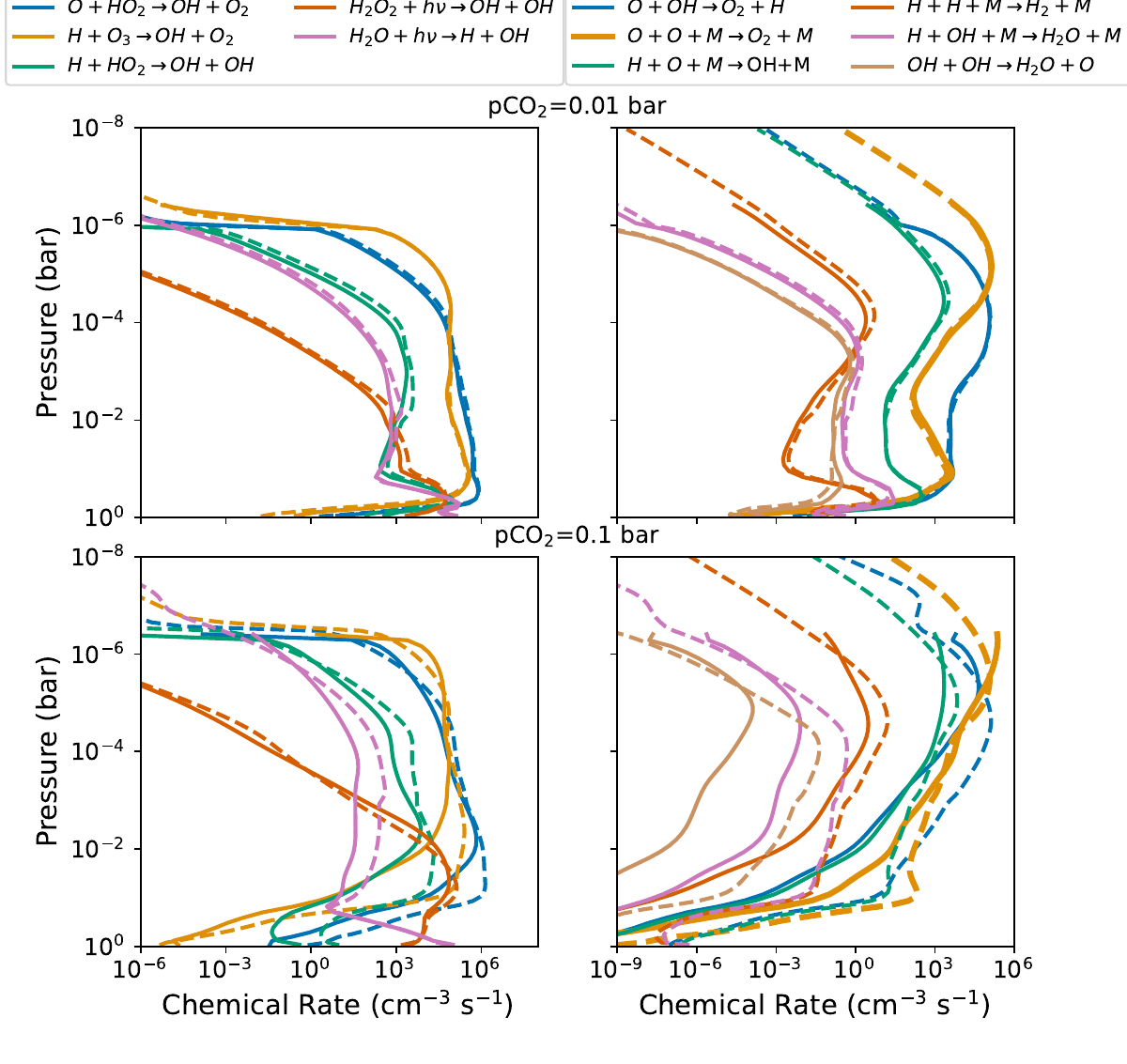}
\caption{Vertically resolved reaction rates of key OH-producing (left column) and radical-radical (right column) reactions, for both truncated (solid lines) and extended (dashed lines) model atmospheres. The top row shows pCO$_2=0.01$ bar, for which both model tops return the same results, while the bottom shows pCO$_2=0.1$ bar, for which CO/O$_2$ runaway has begun in the truncated model top (but for which an NUV-attenuating ozone layer has not yet formed). CHO chemistry only (S, N excluded). The key reaction O+O+M$\rightarrow$ O$_2$+M in the right column is highlighted with a thicker line. Concentration of O production in a single model bin in the truncated atmosphere leads to more intense O-O reactions, which are chain termination reactions which remove radicals from the atmosphere and impede the radical-driven recombination of CO and O$_2$.}
\label{fig:runaway_progression_vertrxn}
\end{figure}

Based on our pCO$_2$=0.01-0.1 bar modeling, we attribute the influence of $\lambda_{crit}$ and $z_{max}$ on CO/O$_2$ runaway to their control on model resolution of the CO$_2$ photolysis peak\footnote{Specifically, the channel $CO_2+h\nu\rightarrow CO + O(^1D)$}. M-dwarf stars emit proportionately more of their UV radiation at far-UV (FUV) wavelengths compared to the Sun. CO$_2$ is a strong absorber of FUV light, and at high CO$_2$ abundances, this FUV radiation is absorbed high in the atmosphere (Appendix~\ref{sec:ap_photodissoc}). Terminating a high-CO$_2$ atmosphere at 54 km (0.34 $\mu$bar for our baseline 0.9 bar CO$_2$, 0.1 bar N$_2$ atmosphere) results in a failure to resolve the distribution of CO$_2$ photolysis in the upper atmosphere with altitude (Figure~\ref{fig:photolysis_peak}). The failure to resolve the CO$_2$ FUV photolysis peak means that CO$_2$ photolysis, and therefore ultimately O production, is artificially confined to the topmost model grid layer. This confinement means that the O atoms which would normally be distributed across a wide altitude range are instead artificially sequestered into a single altitude bin, facilitating their mutual reaction via $\textrm{O+O+M}\rightarrow \textrm{O}_2+\textrm{M}$ (Figure~\ref{fig:runaway_progression_vertrxn}). This is a chain termination reaction, which removes reactive power in the form of radicals from the atmosphere \citep{Grenfell2018}. In particular, the mixing ratio of OH ($X_{OH}$) is  strongly suppressed in truncated atmospheres, and since OH is the main catalyst driving CO$_2$ recombination \citep{Harman2018}, it is unsurprising that as $X_{OH}$ falls, pCO and pO$_2$ rise (Figure~\ref{fig:runaway_progression_guts}; Appendix~\ref{sec:ap_catalyticcycles}). That the CO/O$_2$ runaway is fundamentally driven by OH is signaled by the fact that CO increases before O$_2$ as pCO$_2$ increases. This is because OH is the only effective photochemical sink on CO \citep{Kasting1990}, meaning variations in $X_{OH}$ translate immediately into variations in pCO, while O$_2$'s sinks are more diverse and its response to $X_{OH}$ more indirect. In this sense, CO is the ``canary in the coal mine'', signaling variations in $X_{OH}$ in advance of O$_2$ \citep{Schwieterman2016}.

We conducted additional numerical experiments to confirm our explanation that CO/O$_2$ runaway is driven by confinement of the O produced by CO$_2$ photolysis to a single model layer. If our explanation is correct, then more efficient vertical transport (higher $K_{z}(z)$) should prevent or delay the photochemical runaway, because more of the O produced in the topmost layer of the atmosphere will be transported to lower altitudes instead of reacting with other O radicals in chain termination events. Indeed, increasing $K_{z}(z)$ to $K_z{z}=10^{3}K_{z,0}(z)$, where $K_{z,0}(z)$ is the eddy diffusion profile in our baseline scenario (Figure~\ref{fig:t_p_eddy}) suppresses pO$_2$ and pCO in our baseline scenario (Table~\ref{tab:pco_po2_modelruns}, Run 13). Similarly, if our explanation is correct, then increasing the number of layers in the model grid by decreasing the size of the altitude layers should retard runaway, because the O produced by photolysis is distributed over multiple atmospheric layers instead of being limited to a single layer where it can more readily react with other O. Indeed, increasing the vertical altitude resolution from 1 km to 0.1 km suppresses pO$_2$ and pCO in our baseline scenario (Table~\ref{tab:pco_po2_modelruns}, Run 12). 

We also investigated if we could induce a CO/O$_2$ runaway in an extended model grid which resolved the CO$_2$ photolysis peak by varying $K_z(z)$ or the altitude resolution, and found that we could not. This suggests that models which resolve the CO$_2$ photodissociation peak are less sensitive to other details of the numerical scheme, as expected for physically accurate solutions. For $z_{max}=100$ km, we were unable to induce a CO/O$_2$ runaway by decreasing $K_z(z)$ by setting $K_z{z}=10^{-3}K_{z,0}(z)$ (Table~\ref{tab:pco_po2_modelruns}, Run 14). We were similarly unable to induce a runaway by decreasing the vertical resolution by a factor of 5 to 5 km, or increasing the vertical resolution by a factor of 5 to 0.2 km. However, we noticed a weak sensitivity of pO$_2$ and pCO to vertical resolution, with pO$_2$ increasing by 40\% and pCO decreasing by 50\% as vertical resolution increased from 1 km to 0.2 km. 

\subsection{Synthetic Spectra}

To explore the observational relevance of CO/O$_2$ runaway, we calculate synthetic transmission spectra of a planet in and out of CO/O$_2$ runaway. We specifically simulate and compare spectra of the model atmospheres corresponding to Runs 4 and 5 of Table~\ref{tab:pco_po2_modelruns}, which are in and out of CO/O$_2$ runaway, respectively. These limited simulations do not constitute a rigorous spectral analysis of the potential observables from abiotic CO$_2$-rich planets orbiting M-dwarf stars. Rather, they are narrowly intended to demonstrate the large amplitude of the spectral features that CO/O$_2$ runaway can generate, and therefore the observational relevance of better understanding this theoretical phenomenon and its triggers.

\begin{figure}[h]
\centering
\includegraphics[width=0.8\linewidth]{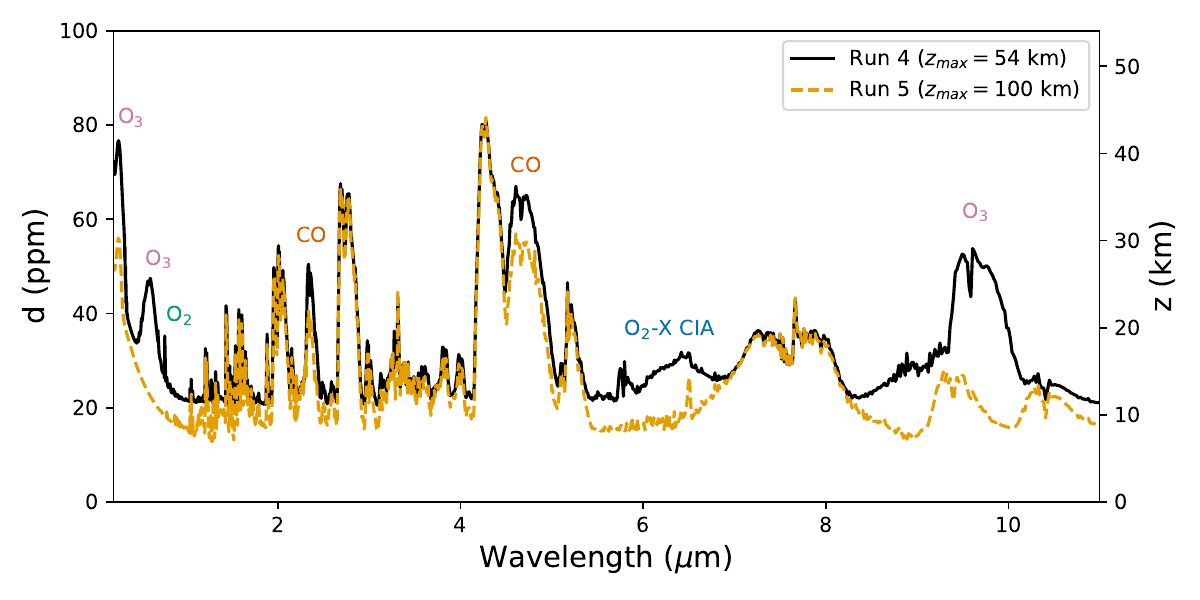}
\caption{Simulated transmission spectra of an Earth-sized planet orbiting a late-M dwarf star with a CO$_2$-dominated atmosphere in and out of CO/O$_2$ runaway, corresponding to Runs 4 and 5 of Table~\ref{tab:pco_po2_modelruns}. Key molecular absorption features that differ between the runaway/non-runaway cases are highlighted. Runaway spectral features are as strong as 50 ppm and the runaway/non-runaway spectra differ by 10s of ppm, which is potentially detectable with JWST.}
\label{fig:simspec}
\end{figure}

Our synthetic spectra show large variations between the runaway (Run 4) and non-runaway (Run 5) atmospheres (Figure~\ref{fig:simspec}). The scale of the spectral features from the runaway atmosphere is comparable to the scale of the spectral features expected from modern Earth-like planets orbiting late M-dwarf stars \citep{Fauchez2020}. The overall amplitude of the transmission spectrum is higher for the runaway case due to the lower mean molecular mass of its atmosphere. The runaway atmosphere displays stronger CO spectral features compared to the non-runaway atmosphere and O$_2$ and O$_3$ spectral features which are absent from the non-runaway atmosphere. The runaway atmosphere displays O$_3$ features as strong as 50 ppm, which are absent from the non-runaway atmosphere. By comparison, JWST has already achieved precision of $<50$ ppm in transmission spectroscopy, with even higher precision potentially achievable with additional observation time given lack of evidence of a photometric noise floor down to a precision of 5 ppm \citep{Lustig-Yaeger2023}. This means that CO/O$_2$ runaway is observationally relevant because it can lead to spectral features that are potentially accessible to JWST. The spectral features due to CO/O$_2$ runaway are also relevant to other facilities such as the Large Interferometer for Exoplanets (LIFE) mission concept \citep{Quanz2022}, and potentially to ground-based facilities as well if observational challenges can be overcome \citep{Hardegree-Ullman2023, Currie2023}.

 % should not be taken as quantitatively definitive predictions of expected spectra of abiotic CO$_2$-rich planets orbiting M-dwarf stars because their underlying photochemical calculations consider CHO chemistry only, ignoring S- and N- bearing species. While such species are not generally expected to be detectable in of themselves on abiotic worlds, they may have higher-order effects on the abundances of their potential higher-order effects on the atmospheric composition and transmission spectrum.

\section{Discussion \label{sec:discuss}}
Our work strengthens the case for O$_2$ as a biosignature gas on conventionally habitable planets orbiting M-dwarf stars. Abundant atmospheric oxygen is the single strongest remotely detectable evidence of life on Earth \citep{Sagan1993}, and O$_2$ detection often motivates design requirements for facilities and surveys to detect life on exoplanets \citep{Rodler2014, Brandt2014, Bixel2021, Checlair2021, Hardegree-Ullman2023}. However, care is required when using O$_2$ as a biosignature because of theoretically-proposed mechanisms for the abiotic accumulation of O$_2$ in the atmospheres of conventionally habitable worlds \citep{Meadows2018}. In particular, it was recently argued that conventionally habitable planets with CO$_2$-rich atmospheres orbiting M-dwarf stars would abiotically generate abundant ($\geq 0.1$ bar) atmospheric oxygen due to photochemical processes \citep{Hu2020}. This regime is highly observationally relevant because (1) CO$_2$-rich atmospheres are expected in the absence of biology, especially for planets orbiting towards the outer edge of the habitable zone \citep{Gaillard2014, Lehmer2020}, and (2) due to observational biases, only planets orbiting M-dwarf stars are accessible to atmospheric characterization with near-term ($\sim10-20$ years) facilities \citep{Shields2016rev}. Notably, TRAPPIST-1e and f, the most observationally favorable potentially habitable exoplanets and which are already targeted by JWST observations, fall into this regime, if conventionally habitable \citep{Wolf2017}. Our results demonstrate that generation of abundant ($\geq 0.1$ bar) O$_2$ on conventionally habitable worlds purely through CO$_2$ photodissociation is unlikely, strengthening the case for O$_2$ as a biosignature. However, photochemical false-positives remain possible for cold, dessicated worlds that are not conventionally habitable \citep{Gao2015}, and escape-driven false-positive mechanisms remain possible for a broad range of worlds \citep{Wordsworth2014, RamirezKaltenegger2014, Luger2015, Tian2015, Wordsworth2016}. 

This paper is the latest in a long series to demonstrate extreme sensitivity of predicted photochemical concentrations of O$_2$ and CO in CO$_2$-rich conventionally habitable planet atmospheres to diverse background model assumptions, including the lightning flash rate, the absorption cross-sections of CO$_2$ and H$_2$O, the rainout rate and atmosphere/ocean redox balance, and now the altitude of the model top \citep{Wen1989, Selsis2002, Segura2007, Harman2015, Harman2018, Ranjan2020}. Such sensitivity is distressing given the desire to use O$_2$ and CO to constrain the presence or absence of biology on exoplanets \citep{Schwieterman2018, Meadows2018, Schwieterman2019, Wogan2020}, for which one desires a robust, assumption-insensitive prediction of photochemical CO or O$_2$. 

Unfortunately, the regime of conventionally habitable planets with thick CO$_2$-rich atmospheres is highly sensitive to background assumptions, because it corresponds to a regime with reduced atmospheric forcing from OH, permitting second-order processes to become important. The trace gas composition of modern Earth's atmosphere is to first order controlled by OH, the ``detergent of the atmosphere'' whose reactivity is the main photochemical control on the abundances of a wide range of trace gases on modern Earth and Mars, and is thought to have done the same for much of their histories \citep{McElroy1972, Parkinson1972, Riedel2008, CatlingKasting2017}. On planets with oxic atmospheres (e.g., modern Earth), OH is robustly produced by the reaction of H$_2$O with O($^{1}$D), sourced primarily from O$_3$ photolysis \citep{Riedel2008}. On anoxic planets with low CO$_2$ abundances (e.g. Neoarchaean Earth) or CO$_2$-dominated but thin atmospheres (modern Mars), OH is robustly produced by the direct photolysis of H$_2$O. This robust production of OH efficiently destroys atmospheric trace gases and triggers catalytic cycles which drive the atmosphere back towards equilibrium \citep{Harman2018}. On anoxic planets with abundant CO$_2$ (e.g., Eoarchaean Earth), this atmospheric forcing is severely inhibited because CO$_2$ largely competes for the same photons as H$_2$O, suppressing OH production \citep{Ranjan2020}. The suppression of the first-order atmospheric control from OH opens the door for second-order effects to become important, explaining the outsized relevance of details of the photochemical scheme in this regime. This effect is amplified on planets orbiting M-dwarf stars, because these stars emit proportionately less light at OH-producing NUV wavelengths \citep{Segura2005}. One might be forgiven for wanting to avoid this regime entirely. However, thick CO$_2$-dominated atmospheres are predicted for Earth early in its history, and are predicted for conventionally habitable planets orbiting near the outer edges of their habitable zones, including the most near-term observationally accessible habitable zone planet, TRAPPIST-1e \citep{Wolf2017, Rugheimer2018, Lehmer2020}. Therefore, this is not a regime that can be neglected in the search for life on exoplanets. 
 
To address this regime properly, we will need to treat the details, which are not yet fully understood in most cases. For example, the NUV cross-sections of CO$_2$ and H$_2$O are still uncertain (\citealt{Wen1989, Ranjan2020}; Broussard et al \textit{in prep}). Importantly, this means that while our results contravene the previously proposed \emph{inevitability} of CO/O$_2$ runaway on CO$_2$-rich M-dwarf exoplanets, they do not completely eliminate the \emph{possibility} of CO/O$_2$ runaway, because it is possible that we have omitted or incorrectly implemented one of these second-order processes. These processes are much less important (and therefore poorly constrained) in a solar system context, but may be highly relevant for M-dwarf planets. In particular, our modeling neglects ion chemistry in the upper atmosphere, which may be an important source of CO and O$_2$ due to thermospheric dissociation of CO$_2$ driven by intense M-dwarf stellar energetic particle fluxes (J. Kasting, private communication, 5/1/2023). It also neglects novel photoreaction channels for H$_2$O, CO$_2$, and CO, which may be relevant in the uppermost atmosphere \citep{An2021, Lo2021, Yang2023}. We conclude that the upper atmosphere is a unexplored potential source of CO/O$_2$ runaway.

While our results show \emph{abundant} ($\geq0.1$ bar) photochemical O$_2$ is unlikely, they nonetheless admit the possibility of \emph{trace} photochemical O$_2$. Our models predict abiotic pO$_2=2\times10^{-5}-4\times10^{-3}$ bar depending on model branch, assumed stellar SED, and assumed lightning flash rate.  Such O$_2$ abundances, while comparable to biotic O$_2$ concentrations on Proterozoic Earth \citep{Harman2015}, do not constitute an observationally relevant biosignature false positive in the near-to-medium-term, because O$_2$ itself is undetectable at such low concentrations -- even much higher, modern Earth-like O$_2$ concentrations (0.2 bar) will be extremely challenging to characterize with either JWST or ELTs \citep{Fauchez2020, Currie2023, Hardegree-Ullman2023}. Detection of trace O$_2$ will likely require advent of purpose-built instruments such as the Habitable Worlds Observatory \citep{Checlair2021, Clery2023}, but HWO will primarily target Sun-like (FGK) stars and will therefore be relatively unaffected by M-dwarf O$_2$ false positives (our work may be relevant to the handful of M-dwarf stars on the HWO target list). More observationally relevant is the false positive potential for O$_3$, which begins to produce IR spectral features in emission at $2\times10^{-3}$ bar for M-dwarf planets \citep{Kozakis2022}. This falls within the range of photochemical pO$_2$ produced in our modeling, meaning that photochemistry may produce observationally relevant O$_3$ false positives on M-dwarf planets. Additionally, there remain parameters we have not explored which may expand pO$_2$ beyond the pO$_2=2\times10^{-5}-4\times10^{-3}$ range spanned by our sensitivity tests to date to model choice, lightning flash rate, and assumed stellar SED, such as volcanic outgassing level, combustion chemistry, and temperature-pressure profile \citep{Selsis2002, Segura2007, Hu2012, Grenfell2018, Harman2022}. Lastly, oceanic chemistry may also impact pO$_2$. For example, the the rate of direct recombination of CO and O$_2$ in marine waters unconstrained, and if efficient may also suppress pO$_2$ and pCO \citep{Harman2015}. We highlight aqueous reactions like O$_2$-CO recombination as priority targets for experimental characterization relevant to understanding exoplanet atmospheres, complementing the acknowledged community priority to characterize gas-phase kinetics \citep{Fortney2016, Fortney2019}. In summary, further theoretical and experimental work is required to establish rigorous limits on predicted trace photochemical O$_2$ and O$_3$ for CO$_2$-rich planets orbiting M-dwarf stars, and understand their context-dependent false positive potential.

Our work illustrates the challenges of projecting atmospheric photochemical models calibrated on Solar System planets to the novel photochemical regimes accessible on exoplanets. A model top of $0.34$ $\mu$bar is capable of reproducing the trace gas composition of modern Earth \citep{Hu2012}. A model top of $0.34$ $\mu$bar is marginally adequate when modeling CO$_2$-rich planets orbiting Sun-like stars, for which the pressure CO$_2$ photodissociation peak pressure is at $\geq1.9$ $\mu$bar (Appendix~\ref{sec:ap_photodissoc}). However, a model top of $0.34~\mu$bar is not adequate when modeling CO$_2$-rich planets orbiting M-dwarf stars, for which the photodissociation peak may be as high as $0.04~\mu$bar (Appendix~\ref{sec:ap_photodissoc}). Care must be taken when projecting to the novel photochemical regimes accessible on exoplanets to ensure that the large diversity of background parameters in photochemical models are correctly calibrated. 

Our work demonstrates the need to consider the upper atmosphere when modeling planets with CO$_2$-rich atmospheres orbiting M-dwarf stars. Most exoplanet photochemical models, including all models currently focused on understanding O$_2$ false positives on temperate terrestrial worlds, focus on the lower atmosphere of the planet (troposphere, stratosphere, mesosphere), because this is the part of the atmosphere most relevant to trace gas detectability and planetary habitability. However, for CO$_2$-rich planets orbiting M-dwarf stars, the CO$_2$ photodissociation peak can be well above the homopause, which corresponds to the thermosphere in the terrestrial atmosphere (Appendix~\ref{sec:ap_homopause}; \citealt{CRC98}). This part of the atmosphere is partially ionized, and ion chemistry is important \citep{Tian2008, Johnstone2018}, which is rarely implemented in the exoplanet photochemistry models being used to study O$_2$ false positives. Similarly, above the homopause, transport is dominated by molecular diffusion, not eddy diffusion, which is not well-captured for all molecules in all models (e.g., MEAC implements molecular diffusion of H and H$_2$ only). Accurately modeling CO$_2$-rich atmospheres orbiting M-dwarf stars will require resolving the thermosphere, and ideally the entire atmosphere from surface to exobase. Efforts towards resolving exoplanet photochemistry in the upper atmosphere alone have been implemented for oxic, steam, and CO$_2$-dominated exoplanet atmospheres \citep{Tian2009b, Garcia-Sage2017, Johnstone2019, Johnstone2020, Nakayama2022}, and from the surface to the upper atmosphere for oxic exoplanet atmospheres \citep{Chen2019, Herbst2019, Cooke2023}. We advocate for the extension of such work to abiotic CO$_2$-dominated atmospheres, which will strongly enhance our understanding of abiotic false positive scenarios for CO and O$_2$.   

Our finds affirm the importance of CO as a discriminant of photochemical O$_2$ production \citep{Schwieterman2016, Wang2016}. CO/O$_2$ runaway is ultimately driven by the suppression of OH, but OH is effectively the \emph{only} photochemical sink on CO in conventionally habitable planet photochemical networks, while numerous pathways can suppress O$_2$. Consequently, when CO/O$_2$ runaways do occur in our model, CO rises before O$_2$. We can concoct scenarios where CO is scrubbed from the atmosphere while O$_2$ rises (e.g, through efficient surface deposition of CO but not O$_2$ into the ocean), but this scenario is unlikely on an abiotic planet based on our current understanding of aqueous chemistry \citep{Kharecha2005, Harman2015, Hu2020}. We therefore continue to advocate for constraining CO abundances in tandem with O$_2$/O$_3$ abundances when seeking to employ O$_2$ as a biosignature, as has also been argued when seeking to employ CH$_4$ as a biosignature \citep{Thompson2022}. Similarly, our work affirms the value of ``capstone" biosignatures to corroborate potentially-ambiguous primary biosignatures like O$_2$ \citep{Leung2022}.

\section{Conclusions \label{sec:conc}}
We have used a multi-model approach to show that accumulation of abundant ($>0.1$ bar) photochemical CO and O$_2$ on conventionally habitable rocky planets with CO$_2$-rich atmospheres orbiting M-dwarf stars (e.g., TRAPPIST-1e, -1f, if habitable) is unlikely, within the limits of current knowledge. An earlier prediction of abundant photochemical CO/O$_2$ on M-dwarf planets was a model artefact, due to a model top that was too low to resolve the CO$_2$ photolysis peak. This model artefact occured because on M-dwarf planets with CO$_2$-rich atmospheres, the CO$_2$ photolysis peak is shifted to very high altitudes, meaning that model tops which are appropriate for solar system planets are too low for the M-dwarf regime. Resolving the CO$_2$ photolysis peak robustly eliminates predictions of abundant ($>0.1$ bar) CO and O$_2$. However, accumulation of trace photochemical O$_2$ ($2\times10^{-4}~\text{bar}<$ pO$_2$ $<$0.1 bar) together with $>1\%$ CO remain a possibility. Such O$_2$ concentrations do not constitute an observationally-relevant biosignature false positive in the near-to-medium term because they are likely too low to be detected directly with JWST, but are high enough that they may drive an observationally-relevant false positive for O$_3$ as a biosignature, and further work is required to address this possibility. Our work demonstrates the need to accurately resolve the upper atmosphere in order to model  O$_2$ false positives on M-dwarf planets, because on such worlds the CO$_2$ photolysis peak can extend into the heterosphere. Overall, our work strengthens the case for O$_2$ as a biosignature gas by reducing its false positive potential, but further work is required to fully understand the context-dependent use of O$_2$ and O$_3$ as biosignature gases.

\begin{acknowledgments}
We thank an anonymous referee for constructive criticism which substantially improved this paper. We thank Roger Yelle and James Kasting for helpful discussions. We thank Thomas Fauchez and Ana Glidden for answers to questions. S.R., E.W.S. and M.L. gratefully acknowledge support by NASA Exoplanets Research Program grant \# 80NSSC22K0235. R.H. was supported by NASA Exoplanets Research Program grant \# 80NM0018F0612. This research has made use of NASA’s Astrophysics Data System. The MEAC input and output files underlying Figures \ref{fig:photolysis_peak}, \ref{fig:runaway_progression_guts}, \ref{fig:runaway_progression_vertrxn}, and Table~\ref{tab:pco_po2_modelruns}, along with the scripts used to analyze them and to conduct the calculations presented in Appendices~\ref{sec:ap_photodissoc} and \ref{sec:ap_homopause}, are publicly available at \url{https://github.com/sukritranjan/co-o2-runaway-revisited-toshare} and via Zenodo \citep{Ranjan2023cZenodo}.
\end{acknowledgments}


\vspace{5mm}
%\facilities{HST(STIS), Swift(XRT and UVOT), AAVSO, CTIO:1.3m, CTIO:1.5m,CXO}


\software{MEAC \citep{Hu2012},  
          Atmos \citep{Arney2016}
          }



\appendix
\counterwithin{figure}{section}

\section{Additional Background: CO$_2$ Photochemical Instability on M-dwarf Planets\label{sec:ap_background}}
In this section, we briefly review the problem of CO$_2$ photochemical instability, with application to M-dwarf planets.

CO$_2$ alone is unstable to photolytic conversion to CO and O$_2$ in conventionally habitable planet atmospheres. CO$_2$ photolyzes efficiently under UV irradiation \citep{ityaksov2008co2}: 
\begin{eqnarray}
    \textrm{CO}_2+\textrm{h}\nu\vert_{<202~\text{nm}} \rightarrow \textrm{CO + O} \\
\end{eqnarray}
The recombination reaction $CO+O+M\rightarrow CO_2+M$ is spin-forbidden and slow, leaving O$_2$ to accumulate via \citep{Krasnopolsky2019}:
\begin{eqnarray}
    \textrm{O + O + M} &\rightarrow \textrm{O}_2\textrm{+M} \\
\end{eqnarray}
However, photodissociation of non-CO$_2$ trace gases can trigger catalytic cycles which recombine CO and O$_2$ back to CO$_2$. For example, on Mars, OH ultimately sourced from H$_2$O photolysis\footnote{While H$_2$O can undergo three-body photolysis as well, this only occurs at EUV wavelengths confined to the upper atmosphere, and initial sensitivity tests indicate no substantial impact on the lower atmosphere \citep{An2021}.} \citep{Kasting1981, Nair1994, Harman2015, Ranjan2020}:
\begin{eqnarray}
    \textrm{H}_2\textrm{O}+\textrm{h}\nu\vert_{<240~\text{nm}}&\rightarrow \textrm{H + OH }\\
\end{eqnarray}
\noindent stabilizes\footnote{On Venus, CO$_2$ recombination is also facilitated by HCl photodissocation; this recombination mechanism is less applicable to conventionally habitable worlds with hydrological cycles, where highly soluble HCl is efficiently scrubbed from the atmosphere by rain \citep{Yung1982, Demore1982, Lightowlers1988}.} CO$_2$, via cycles like \citep{Harman2018}:
\begin{eqnarray}
    \textrm{CO + OH} \rightarrow \textrm{CO}_2+\textrm{H} \\
    \textrm{O}_2\textrm{+H+M} \rightarrow \textrm{HO}_2 + \textrm{M} \\
    \textrm{HO}_2+\textrm{O} \rightarrow \textrm{O}_2+\textrm{OH} \\
\end{eqnarray}

The challenge for M-dwarf exoplanets is that M-dwarfs emit proportionately less near-UV (NUV; $>200$ nm) radiation due to their cooler photospheres, while their emission of far-UV  (FUV; $<200$ nm) remains robust due to their enhanced magnetic activity \citep{Loyd2016}. At habitable temperatures, CO$_2$ known to photoabsorb significantly only at FUV wavelengths \citep{ityaksov2008co2}, while H$_2$O is known to photoabsorb at both FUV and NUV wavelengths \citep{Ranjan2020}. The high FUV-to-NUV emission ratio of M-dwarf stars therefore results in reduced efficacy of the H$_2$O photolysis-driven chemical cycles which recombine CO$_2$ relative to CO$_2$ photolysis, resulting in predictions of enhanced CO and O$_2$ relative to planets orbiting M-dwarf stars. The problem is excacerbated for high-CO$_2$ planets, because abundant CO$_2$ shields H$_2$O from photodissociation, as H$_2$O is restricted to the lower atmosphere by condensation \citep{Harman2015}. If the relative efficacy of H$_2$O photolysis-driven CO$_2$ recombination cycles decreases below a critical point, the atmosphere can enter a state of photochemical runaway, where CO and O$_2$ accumulate to appreciable atmospheric concentrations; this state is termed CO/O$_2$ runaway \citep{Zahnle2008, Harman2015, Hu2020}. The debate in the field has been which conventionally habitable M-dwarf planets, if any, access this CO/O$_2$ runaway atmospheric state \citep{Domagal-Goldman2014, Tian2014, Harman2015, Gao2015, Harman2018, Hu2020}. In the most recent iteration of this debate, \citet{Hu2020} find the CO/O$_2$ runaway regime is robustly accessed for high-CO$_2$ terrestrial planets orbiting M-dwarf stars; in this paper, we test this finding.

%The most obvious cause for a CO/O2 runaway is the lack of reactive catalysts, such as HOx radicals resulting from H2O photolysis. Gao et al. (2015) demonstrate this scenario for very dry CO2-rich planets orbiting M dwarf stars (where high FUV fluxes drive CO$_2$ photolysis while low NUV fluxes arrest its recombination). However, the lack of water vapor bands or other H-bearing gases would be diagnostic of this scenario, and would preclude the habitable planetary environments required for robust biosignature production. The more pernicious scenario is one where abiotic O2 may accumulate in the presence of habitable conditions and their attendant remote observables (i.e., water vapor), complicating biosignature searches and interpretations (see, e.g., [many papers could be cited here]. 

\section{Catalytic Cycles in Truncated vs. Non-Truncated Atmospheres \label{sec:ap_catalyticcycles}}
In this section, we more closely examine the mechanism by which suppressed O abundances in truncated atmospheres leads to decreased production of OH and therefore reduced stability of CO$_2$-rich atmospheres to photolysis. Specifically, we follow \citet{Harman2018} in constructing tables of the main catalytic cycles stabilizing CO$_2$ and their rates in the truncated and untruncated atmospheres with pCO$_2$=0.1 bar (Table~\ref{tab:catalytic_cycles}). The rate of the net catalytic cycle is set by the slowest reaction within the cycle. 

\begin{deluxetable}{ccc}
\tablecaption{Column-Integrated Rates of CO$_2$ Photolysis and Recombination, following \citet{Harman2018}. We compare truncated and nontruncated atmospheres for pCO$_2=0.1$ bar, CHO chemistry only. Cycles 1 and 3 are ultimately drawn from the work of \citet{Stock2012} on the Martian atmosphere. Cycle 2 is identified in this work, where it is relevant to the high-CO$_2$, high-CO atmospheres we consider. The CO$_2$ recombination mechanisms we identify here balance $>90\%$ of CO$_2$photolysis, meaning we have identified the main CO$_2$ recombination mechanisms.  \label{tab:catalytic_cycles}}
\tablewidth{0pt}
\tablehead{\colhead{Reaction} & \colhead{Column-Integrated Rate ($z_{max}=\vert_{P=0.34~\mu}$bar)} & \colhead{Column-Integrated Rate ($z_{max}=100$ km)} \\
& (cm$^{-2}$ s$^{-1}$) & (cm$^{-2}$ s$^{-1}$)}
\startdata
\multicolumn{3}{c}{\textbf{CO$_2$ Photolysis}}\\
$\textrm{CO}_2 + \textrm{h}\nu \rightarrow \textrm{CO + O}$ & 2.1E12 & 3.8E12\\
$\textrm{CO}_2 + \textrm{h}\nu \rightarrow \textrm{CO} + \textrm{O}(^{1}\textrm{D})$ & 2.6E11 & 1.2E11\\
Total CO$_2$ Photolysis & 2.4E12 & 3.9E12\\
\hline
\multicolumn{3}{c}{\textbf{CO$_2$ Recombination Cycle 1}}\\
$\textrm{H} + \textrm{O}_2 +\textrm{M} \rightarrow \textrm{HO}_2+\textrm{M}$ & 3.0E11 & 2.3E12\\
$\textrm{O}+\textrm{HO}_2\rightarrow \textrm{OH}+\textrm{O}_2$ & 1.0E12 & 2.6E12\\
$\textrm{CO+OH}\rightarrow \textrm{CO}_2+\textrm{H}$ & 1.4E12 & 3.5E12\\
Net ($\textrm{CO+O}\rightarrow \textrm{CO}_2$) & 3.0E11 & 2.3E12\\
\hline
\multicolumn{3}{c}{\textbf{CO$_2$ Recombination Cycle 2}}\\
$\textrm{H + CO + M}\rightarrow \textrm{CHO + M}$ & 9.7E11 & 7.4E11\\
$\textrm{CHO + O}_2 \rightarrow \textrm{CO + HO}_2$ & 9.6E11 & 7.1E11\\
$\textrm{O+HO}_2 \rightarrow O\textrm{H + O}_2$ & 1.0E12 & 2.6E12\\
$\textrm{CO+OH}\rightarrow \textrm{CO}_2+\textrm{H}$ & 1.4E12 & 3.5E12\\
Net ($\textrm{CO+O}\rightarrow \textrm{CO}_2$) & 9.6E11 & 7.1E11\\
\hline
\multicolumn{3}{c}{\textbf{CO$_2$ Recombination Cycle 3}}\\
$\textrm{H}_2\textrm{O}_2+\textrm{h}\nu\rightarrow 2\textrm{OH}$ & 9.5E10 & 1.3E11\\
$2\textrm{CO} + 2\textrm{OH} \rightarrow 2\textrm{CO}_2+2\textrm{H}$ & 7.1E11 & 1.8E12\\
$2\textrm{H} + 2\textrm{O}_2+2\textrm{M} \rightarrow 2\textrm{HO}_2 + \textrm{M}$ & 1.5E11 & 1.1E12\\
$2\textrm{HO}_2 \rightarrow \textrm{H}_2\textrm{O}_2+\textrm{O}_2$ & 8.6E10 & 1.0E11\\
Net ($2\textrm{CO+O}_2\rightarrow 2\textrm{CO}_2$) & 8.6E10 & 1.0E11\\
\multicolumn{3}{c}{\textbf{Direct CO$_2$ Recombination}}\\
$\textrm{CO + O + M}\rightarrow \textrm{CO}_2 + \textrm{M}$ & 9.5E11 & 3.7E11\\
\hline
Total CO$_2$ Recombination & 2.4E12 & 3.6E12\\
\multicolumn{3}{l}{(Cycle 1 + Cycle 2 + 2$\times$Cycle 3 + Direct Recomb.)}\\
\enddata
%\tablecomments{Text} 
\end{deluxetable}

\begin{figure}[h]
\centering
\includegraphics[width=0.8\linewidth]{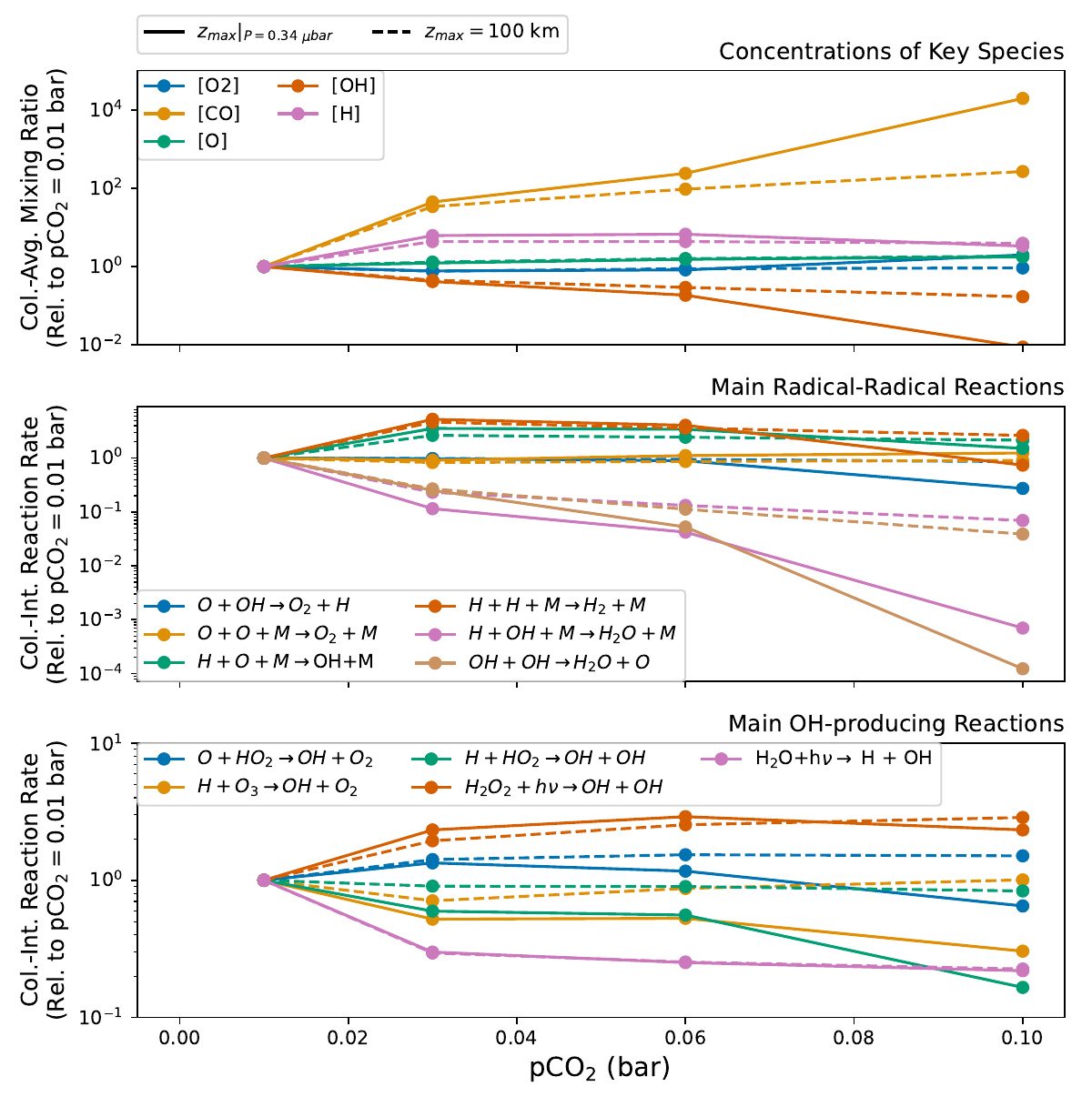}
\caption{A closer look at the mechanics of CO/O$_2$ runaway. Change in column-integrated mixing ratio of key species (top) and column-integrated reaction rates for main radical species (middle) and main OH-producing reactions (bottom) as a function of pCO$_2$, relative to pCO$_2$=0.01 for which both truncated (solid lines) and extended (dashed lines) model calculations produce the same result. Suppression in OH concentrations, likely due to suppression of OH production pathways, underlies CO/O$_2$ runaway.}
\label{fig:runaway_progression_guts}
\end{figure}

The chemical cycles in the C-H-O system are complex and nonlinear, but the catalysis tables lend some insight \citep{Harman2018}. H plays a key role in recombining CO$_2$, because it facilitates generation of intermediate species en route to OH:
\begin{eqnarray}
        \textrm{H+O}_2+\textrm{M}\rightarrow \textrm{HO}_2 + \textrm{M}\\
    \textrm{O}+\textrm{HO}_2\rightarrow \textrm{OH+O}_2\\
    \text{Net: } \textrm{O + H} \rightarrow \textrm{OH}
\end{eqnarray}
\noindent and
\begin{eqnarray}
    \textrm{H+CO+M}\rightarrow \textrm{CHO + M}\\
    \textrm{CHO+O}_2\rightarrow \textrm{CO + HO}_2\\
    \textrm{O+HO}_2\rightarrow \textrm{OH+O}_2\\
    \text{Net: } \textrm{O + H} \rightarrow \textrm{OH}
\end{eqnarray}
However, OH and O are also main sources of H, via:
\begin{eqnarray}
    \textrm{OH + CO}\rightarrow \textrm{CO}_2+\textrm{H}\\
    \textrm{O+OH}\rightarrow \textrm{O}_2+\textrm{H}
\end{eqnarray}
A decrease in O therefore decreases H production, which decreases OH production, which further decreases H production which relies on OH. This is a positive feedback loop, whereby small changes in [O] can be amplified into larger changes in [OH]. We suggest this positive feedback loop as the mechanism whereby the relatively modest changes in [O] driven by the truncated model top (Figure~\ref{fig:runaway_progression_guts}) can be amplified into relatively large changes in [OH], and therefore pCO and pO$_2$.

\section{SED-Driven Changes in Photodissociation Peak Pressure  \label{sec:ap_photodissoc}}
The higher FUV/NUV ratio associated with M-dwarf irradiation leads to lower CO$_2$ photolysis peak pressures (higher CO$_2$ photolysis peak altitudes) relative to Sun-like stars. To illustrate this phenomenon, we derive an estimate of the pressure of peak dissociation for a well-mixed gas $g$ in hydrostatic equilibrium, $P_{peak, g}$, and apply it to CO$_2$ in the atmospheres of CO$_2$-dominated planets orbiting different stars. This calculation is crude; for example, CO$_2$ mixing ratios may decrease with altitude in the uppermost planetary atmosphere due to robust photolysis and diffusive separation, therefore necessitating going deeper into the atmosphere to get the same absorption compared to the well-mixed case. However, the general trends of this calculation give some intuition for why M-dwarf SEDs lead to higher $P_{peak, CO_{2}}$.

The pressure of peak photodissociation of a gas $g$ occurs when the UV optical depth of that gas $\tau_{g, UV}$ (measured traveling from space downwards normal to the planet surface) satisfies \citep{CatlingKasting2017}:
\begin{equation}
    \frac{\tau_{g, UV}}{\cos(\theta_0)}=1 \label{eqn:peak_photo_def}
\end{equation}
\noindent where $\theta_0$ is the stellar angle of incidence.  For an atmosphere in hydrostatic equilibrium \citep{CatlingKasting2017},
\begin{equation}
    P=\bar{\mu}gN \label{eqn:atm_column_mass}
\end{equation}
\noindent where $P$ is the pressure, $\mu$ is the mean molecular weight of the atmosphere, $g$ is the acceleration due to gravity, and $N$ is the number column density of the atmosphere above pressure $P$. Since $\tau_{UV,g}=N_g\overline{\sigma_{UV,g}}$, where $\overline{\sigma_{UV,g}}$ is the mean UV photolysis cross section of $g$, for a well mixed gas with molar concentration $r_g$ (i.e., $N_g=r_gN$), we can substitute Equation~\ref{eqn:atm_column_mass} into Equation~\ref{eqn:peak_photo_def} to write

\begin{equation}
    P_{peak, g}=\frac{\cos(\theta_0)\bar{\mu}g}{r_{g}\overline{\sigma_{UV,g}}} \label{eqn:peak_phot_pressure}
\end{equation}

We apply Equation~\ref{eqn:peak_phot_pressure} to estimate the peak CO$_2$ photolysis pressure for planets orbiting different kinds of stars for our CO$_2$-dominated baseline scenario. For our CO$_2$-dominated baseline scenario, $r_{CO_{2}}=0.9$, $g=g_\earth$, $\mu=42.4$ amu, and we have chosen $\theta_0=57.3^\circ$ \citep{Zahnle2008}. To calculate $\overline{\sigma_{UV,CO_{2}}}$, we calculate the mean CO$_2$ cross-section weighted by the stellar number flux $F_{*}$:

\begin{equation}
    \overline{\sigma_{UV,g}}=\frac{\int_{\lambda_{0}}^{\lambda_{1}} d\lambda  F_{*}(\lambda) \sigma_{CO_{2}}(\lambda)}{\int_{\lambda_{0}}^{\lambda_{1}} d\lambda F_{*}(\lambda)}
\end{equation}

\begin{figure}[h]
\centering
\includegraphics[width=0.8\linewidth]{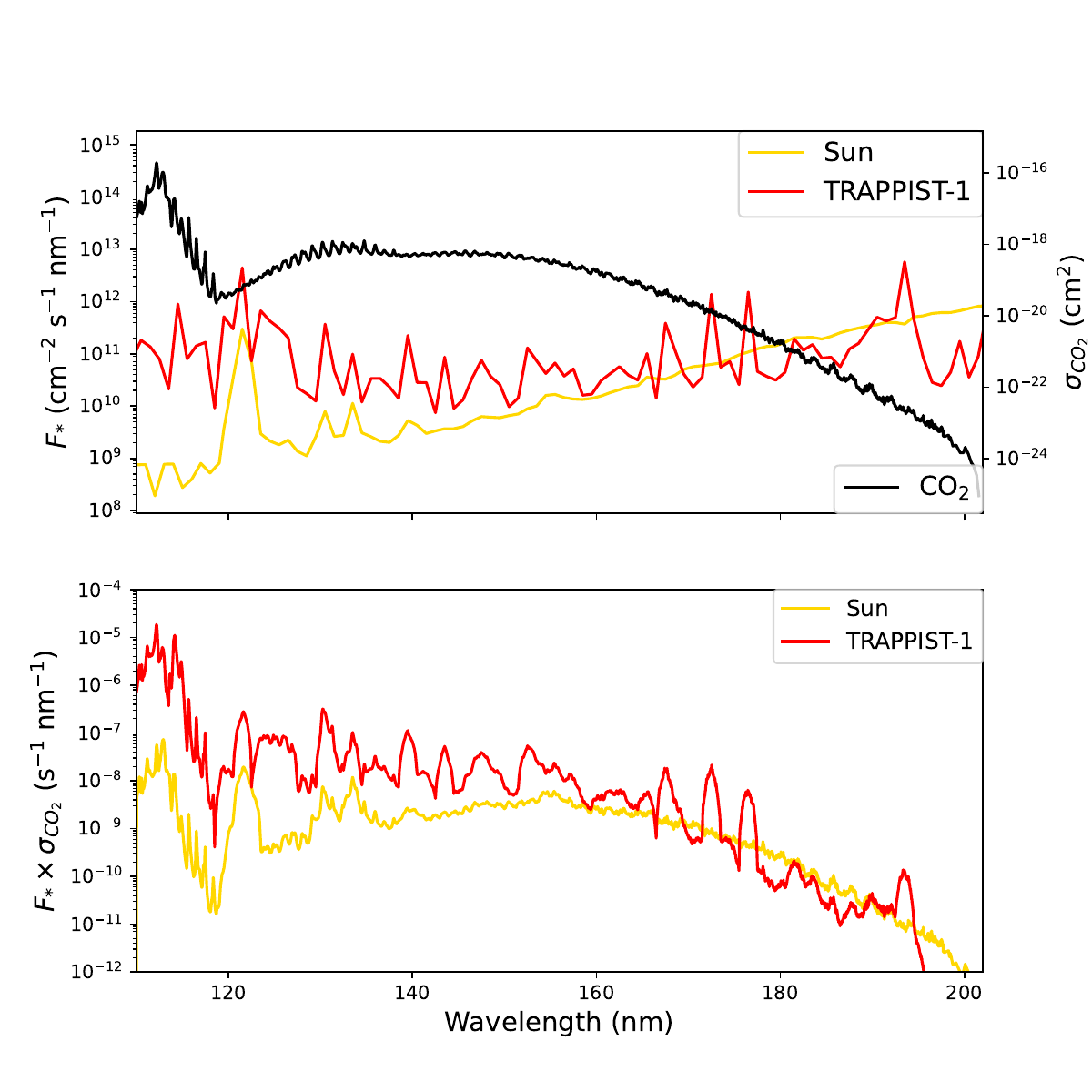}
\caption{Top: Top-of-atmosphere number flux from the Sun and TRAPPIST-1 (left y-axis). Also co-plotted are the photolysis cross-sections of CO$_2$ (right abscissa). Bottom: CO$_2$ cross-sections weighted by the stellar emission of TRAPPIST-1 and the Sun. The bluer spectrum of TRAPPIST-1 increases the weight of shortwave $\sigma_{CO_{2}}$, which is larger, driving an increase in $\overline{\sigma_{UV,g}}$.}
\label{fig:photopeakcalc}
\end{figure}

We calculate $\overline{\sigma_{UV,CO_{2}}}$ and $P_{peak, CO_{2}}$ for Solar irradiation, TRAPPIST-1 irradiation (\citealt{Peacock2019a} model 1a) with $\lambda_{0}=110$ nm, and TRAPPIST-1 irradiation with $\lambda_{0}=120$ nm (Table~\ref{tab:photopeaks}). $\sigma_{CO2}(\lambda)$ and $F_{*}(\lambda)$ for both scenarios are illustrated in Figure~\ref{fig:photopeakcalc}. In all cases, $\lambda_1=202$ nm, corresponding to the end of detected CO$_2$ absorption at room temperature \citep{ityaksov2008co2}. The higher FUV/NUV ratio of TRAPPIST-1 irradiation increases $\overline{\sigma_{UV,g}}$ by  $50\times$ relative to Solar irradiation, decreasing $P_{peak, g}$ proportionately. This means that to resolve the CO$_2$ photolysis peak correctly for planets orbiting M-dwarf stars, it is necessary to extend the model grid to much lower pressures compared to planets orbiting Sunlike stars. 

\begin{deluxetable}{cccc}
\tablecaption{Estimates of $\overline{\sigma_{UV,g}}$ and $P_{peak, CO_{2}}$ for different stellar irradiation. The higher FUV/NUV ratio of M-dwarf stars (here represented by the late-M endmember TRAPPIST-1) lead to higher $\overline{\sigma_{UV,g}}$ and lower $P_{peak, CO_{2}}$ relative to Sun-like stars. \label{tab:photopeaks}}
\tablewidth{0pt}
\tablehead{\colhead{Star} & \colhead{$\lambda_0$} & \colhead{$\overline{\sigma_{UV,g}}$} & \colhead{$P_{peak, CO_{2}}$} \\
& (nm) & cm$^2$& ($\mu$bar)}
\startdata
Sun & 110 & $2\times10^{-20}$ & 2\\
TRAPPIST-1 & 120 & $7\times10^{-20}$ & 0.6\\
TRAPPIST-1 & 110 & $9\times10^{-19}$ & 0.04\\
\enddata
%\tablecomments{Text} 
\end{deluxetable}

In this calculation, we have approximated CO$_2$ as well-mixed. This is a good assumption for the lower atmosphere, but our extended atmospheric simulations extend into the heterosphere where CO$_2$ is not well-mixed, and its abundance decreases with altitude. Therefore, our calculation, which assumes constant CO$_2$ abundance with altitude, formally constitutes a lower bound on $P_{peak, CO_{2}}$. Nevertheless, it illustrates the general mechanism by which $P_{peak, CO_{2}}$ is decreased on planets orbiting M-dwarf stars, i.e. the higher FUV/NUV ratio increasing $\overline{\sigma_{UV,g}}$. 

\section{Homopause Estimates\label{sec:ap_homopause}}
In this Appendix, we estimate the homopause pressure for CO$_2$-dominated temperate terrestrial planets to demonstrate that the CO$_2$ photodissociation peak can extend into the heterosphere for M-dwarf planets, demonstrating the need to resolve the upper atmosphere when modeling such worlds. 

The homopause pressure is the pressure level in the atmosphere when 
\begin{equation}
    K_z(z)=D_{12}(T, n) \label{eqn:homopausedef}
\end{equation}

\noindent where $D_{12}(T, n)$ is the diffusion coefficient of gas $g_1$ diffusing through a background gas $g_2$ at temperature $T$ and atmospheric number density $n$. $D_{12}(T, n)$ takes general form \citep{Banks1973}:
\begin{equation}
    D_{12}(T, n)=\frac{A_{12}T^{s_{12}}}{n} \label{eqn:diffcoeffgenform}
\end{equation}
\noindent where $A_{12}$ and $s_{12}$ are specific to each ($g_1$, $g_2$) pair. Combining equations~\ref{eqn:homopausedef} and \ref{eqn:diffcoeffgenform} with the ideal gas law ($P=nkT$), we can write
\begin{equation}
    P_{homo}=\frac{k A_{12} T_{homo}^{1+s_{12}}}{K_{z, homo}} \label{eqn:homopause}
\end{equation}

In Earth's upper atmosphere near the homopause, $K_z(z)=4\times10^{5}-2\times10^{6}$ cm$^{2}$ s$^{-1}$, depending on which tracer is used \citep{Hunten1975, Swenson2019}. We adopt an Earth upper atmosphere $K_{z,homo,\earth}=1\times10^{6}$ cm$^{2}$ s$^{-1}$, which may be scaled to $K_{z,homo, CO_{2}-dominated}=\frac{29.0}{42.4}(1\times10^{6}~\text{cm}^{-2}~\text{s}^{-1}\text{)}=7\times10^{5}~\text{cm}^{-2}~\text{s}^{-1}$ for our CO$_2$-dominated benchmark scenario \citep{Hu2012}. Further, in the upper atmosphere, $T=175$K in our baseline scenario, meaning we can set $T_{homo}=175$ K. With these values, we evaluate Equation~\ref{eqn:homopause} for Ar, H$_2$, and H diffusing through CO$_2$ and find homopause pressures of $0.07~\mu$bar, $0.4~\mu$bar, and $0.6~\mu$bar, respectively (Table~\ref{tab:homopauses}). For comparison, the photolysis peak pressure is $0.1~\mu$bar for pCO$_2$=0.1 bar (Figure~\ref{fig:photolysis_peak}), and $0.01~\mu$bar for pCO$_2$=0.9 bar as in our baseline scenario. This means that to accurately model the photochemistry of CO$_2$-rich planets orbiting M-dwarf stars, it is necessary to extend model grids well into the heterosphere. We have done so here, but our calculation is not accurate because it does not accurately represent vertical transport in this part of the atmosphere, especially molecular diffusion, and because it does not include the ion chemistry important in the thermosphere and above. We advocate for the construction of exoplanet photochemistry models that resolve the atmosphere from surface to exobase to accurately treat the important, observationally-relevant problem of CO$_2$-rich atmospheres orbiting M-dwarf stars, and especially their propensity to accumulate photochemical CO and O$_2$. 

\begin{deluxetable}{cccccc}
\tablecaption{Homopause estimates for different gases in our CO$_2$-dominated baseline planetary scenario. The CO$_2$ photodissociation peak for CO$_2$-dominated atmospheres lie in the heterosphere. \label{tab:homopauses}}
\tablewidth{0pt}
\tablehead{\colhead{Gas} & \colhead{Background Gas} & \colhead{$A_{12}$} & \colhead{$s_{12}$} & $P_{homo}$ ($\mu$bar) & Refs.}
\startdata
Ar & CO$_2$ & 7.16E16 & 0.646 & 0.07& \citet{Marrero1972}\\
H2 & CO$_2$ &  2.15E17 & 0.75 & 0.4 &\citet{Marrero1972, Hu2012}\\
H & CO$_2$ &  3.87E17 & 0.75 & 0.6 & \citet{Marrero1972, Zahnle2008, Hu2012}\\
\enddata
%\tablecomments{Text} 
\end{deluxetable}

\section{Detailed Planet Scenario\label{sec:ap_detailed_scenario}}
\begin{figure}[h]
\centering
\includegraphics[width=0.8\linewidth]{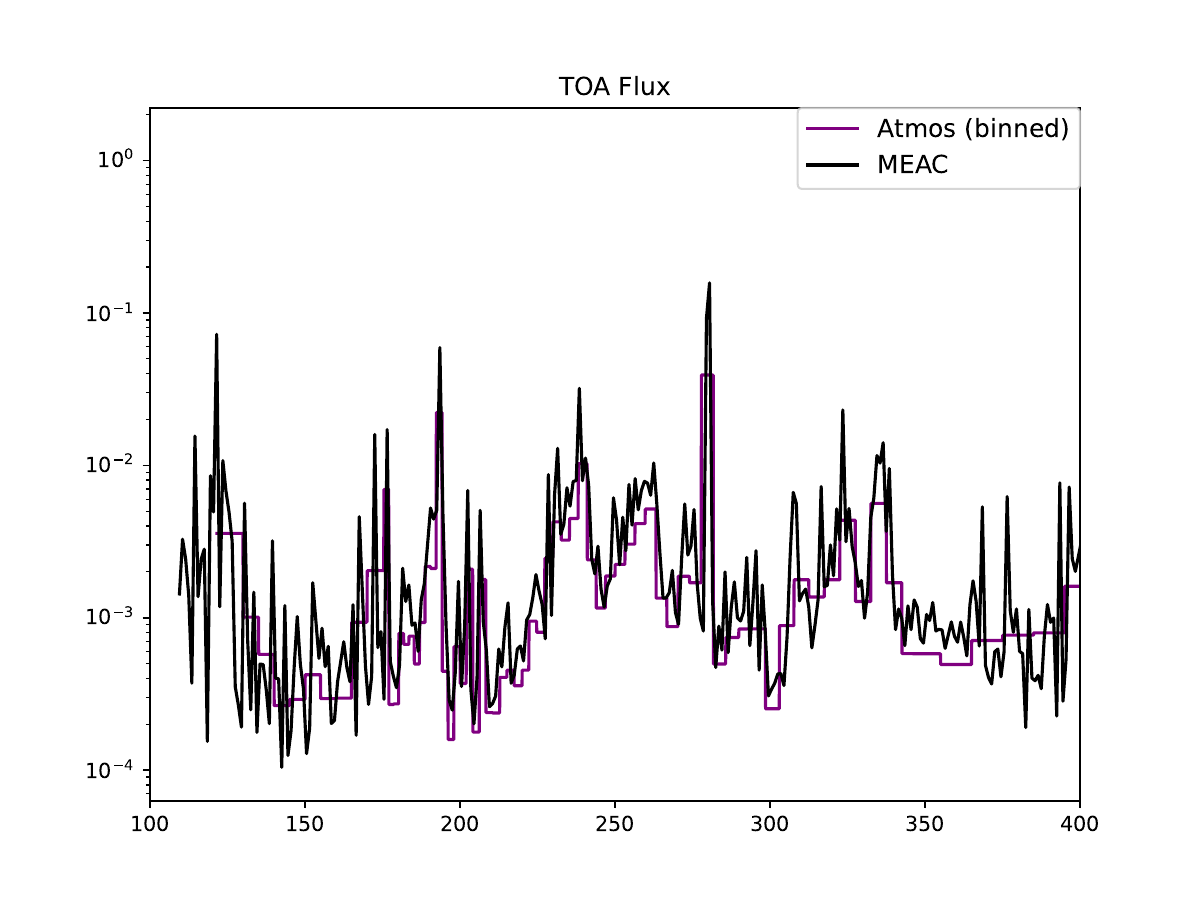}
\caption{TRAPPIST-1 spectra employed by MEAC and Atmos in this study.}
\label{fig:stellar_inputs}
\end{figure}

\begin{figure}[h]
\centering
\includegraphics[width=0.8\linewidth]{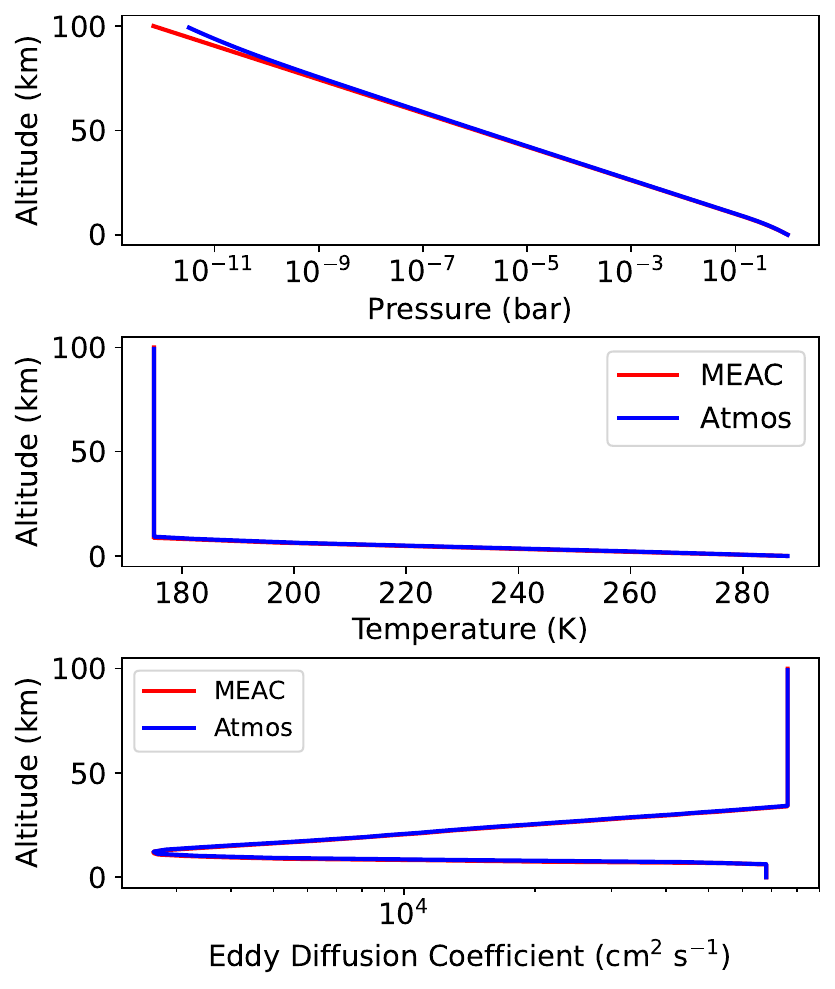}
\caption{T-P and Eddy diffusion profiles employed by MEAC and Atmos in this study.}
\label{fig:t_p_eddy}
\end{figure}

In this Appendix, we give more details of our planetary scenario to facilitate reproduction of our work. Table~\ref{tab:detailed_parameters} summarizes the further details on our general planetary scenario and model set-ups. Table~\ref{tab:species_bcs} presents the detailed species-by-species boundary conditions employed in our simulations. Figure \ref{fig:stellar_inputs} shows the stellar SEDs assumed in this work. Figures~\ref{fig:t_p_eddy} shows the baseline atmospheric structure profiles ($T(z)$, $P(z)$, and $K_z(z)$) assumed in this work. In constructing the atmospheric structure profiles for the variable pCO$_2$ runs shown in Figures~\ref{fig:photolysis_peak}, \ref{fig:runaway_progression_guts}, and \ref{fig:runaway_progression_vertrxn}, we followed \citet{Hu2012}, i.e. assuming dry adiabatic evolution to a 175K stratosphere (with thermodynamic parameters drawn from \citealt{PPC}) and scaling modern Earth's eddy diffusion profile by bulk atmospheric mean molecular mass. 

\begin{deluxetable}{ccc}
%\tablenum{2}
\tablecaption{Simulation Parameters For Planetary Scenario.\label{tab:detailed_parameters}}
\tablewidth{0pt}
\tablehead{\colhead{Scenario Parameter} & \colhead{MEAC} & \colhead{ATMOS}}
\startdata
Model Reference & \citet{Hu2012} & \citet{Arney2016}\\
Reaction Network & As in \citet{Ranjan2022b}$^\dagger$ & Archean Scenario \\
& (Excludes, C$_{>2}$-chem) & (includes \citet{Ranjan2020} updates)\\
\hline
Stellar type & TRAPPIST-1 \citep{Peacock2019a} & TRAPPIST-1 \citep{Peacock2019a}\\
Stellar Constant Relative to Modern Earth & 0.592 & 0.592\\
Planet size & 1 M$_{\earth}$, 1 R$_{\earth}$ &  1 M$_{\earth}$, 1 R$_{\earth}$ \\
Surface albedo &  0. & 0.25 \\ 
Major atmospheric components & 0.9 bar CO$_2$, 0.1 bar\ N$_2^\star$& 0.9 bar CO$_2$, 0.1 bar\ N$_2^\star$ \\
Surface temperature ($z=0$ km) & 288K & 288K \\
Surface $r_{H_{2}O}$ (lowest atmospheric bin) &  0.01 &  0.01\\
Eddy Diffusion Profile &  See Figure~\ref{fig:t_p_eddy} &  See Figure~\ref{fig:t_p_eddy}\\
Temperature-Pressure Profile &  See Figure~\ref{fig:t_p_eddy} &  See Figure~\ref{fig:t_p_eddy}\\
Vertical Resolution & 1 km & 0.5 km\\
\hline
Rainout & Earthlike; rainout turned off for   & Earthlike (all species) \\
& H$_2$, CO, CH$_4$, NH$_3$, N$_2$, C$_2$H$_2$, & \\
& C$_2$H$_4$, C$_2$H$_6$, and O$_2$ to simulate & \\
& saturated ocean on abiotic planet& \\
Lightning & Simulated via $\phi_{CO}(z=0)=\phi_{NO}(z=0)=1\times10^{8}$& On \\
Global Redox Conservation & No & No \\
\enddata
\tablecomments{$^\star$ In MEAC, pN$_2$ is fixed. In Atmos, pN$_2$ is adjusted to maintain dry $P=1$ bar.} 
\tablecomments{$^\dagger$. Default full-chemistry CHOSN model runs contain 86 species linked by 734 reactions. For model runs where we exclude HNO$_4$ and N$_2$O$_5$, our network contains 84 species linked by 723 reactions. For model runs where we exclude S and N chemistry entirely, we set the initial concentrations and emission fluxes of S- and N-containing species to 0, but otherwise retain the full 734-reaction photochemical network. } 
\end{deluxetable}

 We draw our TRAPPIST-1 spectrum from \citet{Peacock2019a}, model \#1A. This spectrum is binned to 1-nm resolution for use in MEAC, and to the standard Atmos model 750 point grid for use in Atmos \citep{lincowski2018evolved}. %We draw our irradiation spectrum for Sunlike stars from \citet{Hu2012}, who in turn synthesized it from the Air Mass Zero reference spectrum produced by the American Society For Testing and Materials\footnote{https://www.nrel.gov/grid/solar-resource/spectra.html} ($\geq119.5$ nm) and from the average quiet Sun emission spectrum of \citet{Curdt2004} ($<119.5$ nm).
 
\startlongtable
\begin{deluxetable}{ccccccc}
\tablecaption{Species choice and treatment for the photochemical models used in this study.\label{tab:species_bcs}}
\tablewidth{0pt}
\tablehead{
\colhead{Species} & \colhead{Type} &  & \colhead{Surface Flux} & \colhead{Surface Mixing Ratio} & \colhead{Dry Deposition Velocity} & \colhead{TOA Flux}\\
  & \colhead{MEAC}& \colhead{ATMOS} & \colhead{(cm$^{-2}$ s$^{-1}$)} & \emph{(relative to CO$_2$+N$_2$)} & \colhead{(cm s$^{-1}$)} & \colhead{(cm$^{-2}$ s$^{-1}$)}
 }
\startdata
CO$_2$ & X & X & -- & 0.9 & 0 & 0 \\
N$_2$ & C & C & -- & 0.1 & 0 & 0 \\
H$_2$O & X & X & -- & 0.01 & 0 & 0 \\
\hline
H$_2$ & X & X & $3\times10^{10}$ & -- & 0 & Diffusion-limited  \\
CO & X & X & $1\times10^{8}$ & -- & $1\times10^{-8}$ & 0 \\
CH$_4$ & X & X & $3\times10^{8}$ & -- & 0 & 0 \\
SO$_2$ & X & X & $3\times10^9$ & -- & 1 & 0 \\
H$_2$S & X & X & $3\times10^{8}$ & -- & 0.015 & 0 \\
\hline
H & X & X & 0 & -- & 1 & Diffusion-limited \\
\hline
O & X & X & 0 & -- & 1 & 0 \\
O(1D) & X & F & 0 & -- & 0 & 0 \\
O$_2$ & X & X & 0 & -- & 0 & 0 \\
O$_3$ & X & X & 0 & -- & 0.4 & 0 \\
OH & X & X  & X & -- & 1 & 0 \\
HO$_2$ & X & X & 0 & -- & 1 & 0 \\
H$_2$O$_2$ & X & X & 0 & -- & 0.5 & 0 \\
CH$_2$O & X & X & 0 & -- & 0.1 & 0 \\
CHO & X & X & 0 & -- & 0.1 & 0 \\
C & X & F & 0 & -- & 0 & 0 \\
CH & X & X & 0 & -- & 0 & 0 \\
CH$_2$ & X & X & 0 & -- & 0 & 0 \\
${}^{1}$CH$_2$ & X & F & 0 & -- & 0 & 0 \\
${}^{3}$CH$_2$ & X & X & 0 & -- & 0 & 0 \\
CH$_3$ & X & X & 0 & -- & 0 & 0 \\
CH$_3$O & X & X & 0 & -- & 0.1 & 0 \\
CH$_4$O & X & - & 0 & -- & 0.1 & 0 \\
CHO$_2$ & X & - & 0 & -- & 0.1 & 0 \\
CH$_2$O$_2$ & X & - & 0 & -- & 0.1 & 0 \\
CH$_3$O$_2$ & X & X & 0 & -- & 0 & 0 \\
CH$_4$O$_2$ & X & - & 0 & -- & 0.1 & 0 \\
C$_2$ & X & X & 0 & -- & 0 & 0 \\
C$_2$H & X & X & 0 & -- & 0 & 0 \\
C$_2$H$_2$ & ? & X & 0 & -- & 0 & 0 \\
C$_2$H$_3$ & ? & X & 0 & -- & 0 & 0 \\
C$_2$H$_4$ & ? & X & 0 & -- & 0 & 0 \\
C$_2$H$_5$ & X & X & 0 & -- & 0 & 0 \\
C$_2$H$_6$ & X & X & 0 & -- & $1\times10^{-5}$ & 0 \\
C$_2$HO & X & - & 0 & -- & 0 & 0 \\
C$_2$H$_2$O & X & - & 0 & -- & 0.1 & 0 \\
C$_2$H$_3$O & X & - & 0 & -- & 0.1 & 0 \\
C$_2$H$_4$O & X & - & 0 & -- & 0.1 & 0 \\
C$_2$H$_5$O & X & - & 0 & -- & 0.1 & 0 \\
S & X & X & 0 & -- & 0 & 0 \\
S$_2$ & X & X & 0 & -- & 0 & 0 \\
S$_3$ & X & F & 0 & -- & 0 & 0 \\
S$_4$ & X & F & 0 & -- & 0 & 0 \\
SO & X & X & 0 & -- & 0 & 0 \\
${}^1$SO$_2$ & X & F & 0 & -- & 0 & 0 \\
${}^3$SO$_2$ & X & F & 0 & -- & 0 & 0 \\
SO$_3$ & X & X & 0 & -- & 1 & 0 \\
HS & X & X & 0 & -- & 0 & 0 \\
HSO & X & X & 0 & -- & 0 & 0 \\
HSO$_2$ & X & - & 0 & -- & 0 & 0 \\
HSO$_3$ & X & F & 0 & -- & 0.1 & 0 \\
H$_2$SO$_4$ & X & X & 0 & -- & 1 & 0 \\
H$_2$SO$_4$(A) & A & A & 0 & -- & 0.2 & 0 \\
S$_8$ & X & - & 0 & -- & 0 & 0 \\
S$_8$(A) & A & A & 0 & -- & 0.2 & 0 \\
OCS	& X	& X & 0 & -- & 0.01 &	0 \\
CS	& X	& X &  0 & -- & 0.01 &	0 \\
CH$_3$S & X	& - & 0 & -- & 0.01 &	0 \\
CH$_4$S & X	& - & 0 & -- & 0.01 &	0 \\
N & X & X & 0 & -- & 0 & 0\\
NH$_3$ & X & - & 0 & -- & 1 & 0\\
NH$_2$ & X & - & 0 & -- & 0 & 0\\
NH & X & - & 0 & -- & 0 & 0\\
N$_2$O & X & - & 0 & -- & 0 & 0\\
NO & X & X & $1.0\times10^{8}$ & -- & 0.02 & 0\\
NO$_2$ & X & X & 0 & -- & 0.02 & 0\\
NO$_3$ & X & - & 0 & -- & 1 & 0\\
HNO & X & X & 0 & -- & 0 & 0\\
HNO$_2$ & X & F & 0 & -- & 0.5 & 0\\
HNO$_3$ & X & X & 0 & -- & 4 & 0\\
HCN & X & - & 0 & -- & 0.01 & 0\\
CN & X & - & 0 & -- & 0.01 & 0\\
CNO & X & - & 0 & -- & 0 & 0\\
HCNO & X & - & 0 & -- & 0 & 0\\
CH$_3$NO$_2$ & X & - & 0 & -- & 0.01 & 0\\
CH$_3$NO$_3$ & X & - & 0 & -- & 0.01 & 0\\
CH$_5$N & X & - & 0 & -- & 0 & 0\\
C$_2$H$_2$N & X & - & 0 & -- & 0 & 0\\
C$_2$H$_5$N & X & - & 0 & -- & 0 & 0\\
N$_2$H$_2$ & X & - & 0 & -- & 0 & 0\\
N$_2$H$_3$ & X & - & 0 & -- & 0 & 0\\
N$_2$H$_4$ & X & - & 0 & -- & 0 & 0\\
C$_3$H$_2$ & - & X & 0 & -- & 0 & 0\\
C$_3$H$_3$ & - & X & 0 & -- & 0 & 0\\
CH$_3$C$_2$H & - & X & 0 & -- & 0 & 0\\
CH$_2$CCH$_2$ & - & X & 0 & -- & 0 & 0\\
C$_2$H$_5$CHO & - & X & 0 & -- & 0 & 0\\
C$_2$H$_5$CHO & - & X & 0 & -- & 0 & 0\\
C$_3$H$_6$ & - & X & 0 & -- & 0 & 0\\
C$_3$H$_7$ & - & X & 0 & -- & 0 & 0\\
C$_3$H$_8$ & - & X & 0 & -- & 0 & 0\\
C$_2$H$_4$OH & - & X & 0 & -- & 0 & 0\\
C$_2$H$_2$OH & - & X & 0 & -- & 0 & 0\\
C$_2$H$_2$CO & - & X & 0 & -- & 0 & 0\\
C$_2$H$_3$CO & - & X & 0 & -- & 0 & 0\\
C$_2$H$_3$CHO & - & X & 0 & -- & 0 & 0\\
HCS & - & X & 0 & -- & 0 & 0\\
CS$_2$ & - & X & 0 & -- & 0 & 0\\
HCAER1 & - & A & 0 & -- & 0 & 0\\
HCAER2 & - & A & 0 & -- & 0 & 0\\
\enddata
\tablecomments{(1) For species type, for each model, ``X" means the full continuity-diffusion equation is solved for the species; ``F" means it is treated as being in photochemical equilibrium; ``A" means it is an aerosol and falls out of the atmosphere; ``C" means it is treated as chemically inert; and ``--" means that it is not included in that model. Note that boundary conditions like dry deposition velocity are not relevant for Type ``F" species, since transport is not included for such species. The exclusion of a species from a model does not necessarily mean that the model is incapable of simulating the species, but just that it was not included in the atmospheric scenario selected here. For example, following \citet{Hu2012}, the MEAC model was deployed here without N species because the planet scenario selected here precludes reactive N, though it is capable of simulating nitrogenous chemistry. (2) For the bottom boundary condition, either a surface flux is specified, or a surface mixing ratio. N$_2$ is a special case in the Kasting model and in ATMOS; in these models, [N$_2$] is adjusted to set the total dry pressure of the atmosphere to be 1 bar (to account for outgassed species and photochemical intermediates). Consequently, pN$_2\lesssim0.1$ bar in these models. (3) TOA flux refers to the magnitude of outflow at the top-of-the-atmosphere (TOA); hence, a negative number would correspond to an inflow. (4) For model runs where we exclude S and N chemistry entirely, the emission fluxes of S- and N-containing species to 0, but otherwise retain the full 734-reaction photochemical network. } 
\end{deluxetable}

\bibliography{sample631}{}
\bibliographystyle{aasjournal}



\end{document}